\documentclass[11pt]{article}
\usepackage{amsfonts}
\usepackage{amssymb}
\usepackage{graphicx}
\usepackage{amsmath}
\usepackage{amsthm}
\usepackage{color}
\usepackage{multirow}
\usepackage{url}
\usepackage{float}
\usepackage{booktabs} 
\usepackage{makecell}
\usepackage{listings}
\usepackage{subcaption}
\usepackage{algorithm}
\usepackage{enumerate}
\usepackage{tikz}

\usepackage{algpseudocode}
\lstnewenvironment{Rcode}{\lstset{language=R}}{}

\newtheorem{theorem}{Theorem}[section]

\newtheorem{lemma}[theorem]{Lemma}

\newtheorem{remark}[theorem]{Remark}

\definecolor{battleshipgrey}{rgb}{0.52, 0.52, 0.51}
\definecolor{navyblue}{rgb}{0.0, 0.0, 0.5}
\definecolor{arsenic}{rgb}{0.23, 0.27, 0.29}
\definecolor{oldmauve}{rgb}{0.4, 0.19, 0.28}
\usepackage[colorlinks=TRUE]{hyperref}
\hypersetup{
	colorlinks=true,
	linkcolor=navyblue,
	filecolor=blue,      
	urlcolor=oldmauve,
	citecolor=navyblue,
	pdftitle={Overleaf Example},
	pdfpagemode=FullScreen,
}
\usepackage{natbib}
\setcitestyle{authoryear}

\begin{document}
		
	\title{ \bf Semi-functional partial linear regression with measurement error: An approach based on $k$NN estimation}
	\author{Silvia Novo{$^{a,b,c}$}\footnote{Corresponding author email address: \href{snovo@est-econ.uc3m.es}{snovo@est-econ.uc3m.es}} \hspace{2pt} Germ\'{a}n Aneiros{$^{d,e}$} \hspace{2pt} Philippe Vieu{$^f$} \\		
		{\normalsize $^a$ Grupo NICDA, Departamento de Estad\'istica, Universidad Carlos III de Madrid, Madrid, Spain}\\
		{\normalsize $^b$ Instituto Flores de Lemus, Madrid, Spain}\\
		{\normalsize $^c$ UC3M-Santander Big Data Institute, Madrid, Spain}\\
		{\normalsize $^d$ Grupo MODES, Departamento de Matemáticas, Universidade da Coruña, A Coruña, Spain}\\
		{\normalsize $^e$ Centro de Investigaci\'on en Tecnolog\'ias de la Informaci\'on y las Comunicaciones (CITIC), A Coruña, Spain}\\
		{\normalsize $^f$ Institut de Math\'{e}matiques, Universit\'e  Paul Sabatier, Toulouse, France}
	}
	
	\date{}
	\maketitle
	\begin{abstract} This paper focuses on a semiparametric regression model in which the response variable is explained by the sum of two components. One of them is parametric (linear), the corresponding explanatory variable is measured with additive error and its dimension is finite ($p$). The other component models, in a nonparametric way, the effect of a functional variable (infinite dimension) on the response. $k$NN based estimators are proposed for each component, and some asymptotic results are obtained. A simulation study illustrates the behaviour of such estimators for finite sample sizes, while an application to real data shows the usefulness of our proposal. 
	\end{abstract}
	
	
\section{Introduction}
Functional data analysis (FDA) is a branch of statistics that analyses data providing information about curves, surfaces or any other mathematical object varying over a continuum, often time. These curves are defined by some functional form and are known as functional data. Nowadays, with the development of modern technology together with the high storage capacity, it is usual for statisticians to deal with functional data, and modelling this kind of data has become one of the most popular topics in Statistics. See, for instance, \cite{rams05} and \cite{ferv06} for some early monographs on parametric and nonparametric modeling of functional data, respectively, and \cite{anehhv22} for a compendium of recent advances in FDA.

Regression analysis is a statistical tool to model the effect of several covariates on a response variable. In general, the regression function is parametric (usually linear), nonparametric or semiparametric.  
The recent advances in FDA (see \citealt{goiav16}, \citealt{anecfgv19}, \citealt{anehhv22}) show that the infatuation for semiparametric modelling in infinite-dimensional framework is increasing day after day. Semiparametric models enjoy flexibility and interpretability, in opposite of parametric and nonparametric models, respectively. In addition, they attenuate the `curse of dimensionality', which is particularly important when dealing with functional data (therefore, infinite-dimensional data). For references on semiparametric modeling of functional data, see, for instance, \cite{anev06}, \cite{wang_2016} and \cite{novo_2019}, and \cite{ling_2021} for a selected advanced review. 

The Semi-Functional Partial Linear (SFPL) regression model is a semiparametric regression model that has been widely studied in recent years. It is defined by the relationship
\begin{equation}
	\label{eq_model1}
	Y=X^{\top}\pmb{\beta}_0+m\left(\mathcal{X}\right)+\varepsilon ,
\end{equation}
where $\pmb{\beta}_0=\left(\beta_{01},\dots,\beta_{0p}\right)^{\top}\in\mathbb{R}^p$ is a vector of unknown parameters, $m(\cdot)$ is an unknown real-valued operator, the response variable ($Y$) takes values in $\mathbb{R}$, $X$ is an explanatory variable taking values in $\mathbb{R}^p$ and $\mathcal{X}$ is another explanatory variable but of functional nature. In addition, the random error, $\varepsilon$, satisfies
\begin{equation}
	\mathbb{E}\left(\varepsilon | X, \mathcal{X}\right)=0. \label{E}
\end{equation}
Throughout all this paper, we will assume that $\mathcal{X}$ is valued in $S_\mathcal{F} \subset \mathcal{F}$, where $\mathcal{F}$ is some abstract infinite-dimensional semi-metric space whose associated semi-metric is denoted by $d(\cdot,\cdot)$.

Model (\ref{eq_model1}) was introduced in \cite{anev06} as an extension of the non-functional partial linear regression model proposed in \cite{eng86} and studied in \cite{spe88}, \cite{rob88}, \cite{gao95} and \cite{ane04} among others; see also the monograph by \cite{har00}. Specifically, those papers dealt with the particular case of model (\ref{eq_model1}) where the covariate $\mathcal{X}$ takes values in $\mathbb{R}^q$ ($1\leq q < \infty $). Now, focusing on the setting of functional $\mathcal{X}$, some theoretical studies on model (\ref{eq_model1}) can be seen in \cite{anev06} (asymptotic distribution and rates of convergence), \cite{sha14} (bandwidth selection), \cite{ane15} (variable selection) and \cite{lin19} (asymptotic distribution and rate of convergence when the response is missing at random), among others.

This paper focuses on the estimation of $\pmb{\beta}_0$ and $m(\cdot)$ in the SFPL model (\ref{eq_model1}) when the covariate $X$ is measured with error; that is, when instead of observing $X$ one observes
\begin{equation}
	\label{EM}
	W=X+U,
\end{equation}
where the measurement error, $U$, has mean zero and covariance matrix $\pmb{\Sigma}_{uu}.$ In addition, $U$ is independent of $(X,\mathcal{X},Y)$. As usual (\citealt{lia99}, \citealt{wei20}, \citealt{zhu20}), 
we will assume that $\pmb{\Sigma}_{uu}$ is known (see Remark \ref{unknown} for the case of unknown $\pmb{\Sigma}_{uu}$). Combination of (\ref{eq_model1}) and (\ref{EM}) gives the so-called SFPL Measurement Error (SFPLME) model. 

In the particular case of non-funcional $\mathcal{X}$,  
model (\ref{eq_model1})-(\ref{EM}) was introduced in \cite{lia99}. They proposed kernel-based estimators for both $\pmb{\beta}_0$ and $m(\cdot)$, as well as for the error variance (say $\sigma^2$); they also obtained the limit distributions of the estimators of $\pmb{\beta}_0$ and $\sigma^2$, and the asymptotic bias and asymptotic variance of the estimator of $m(\chi)$ ($\chi$ denotes an observation of the random variable $\mathcal{X}$). In addition, \cite{wei20} derived the law of the iterated
logarithm for the estimators of $\pmb{\beta}_0$ and $\sigma^2$ proposed by \cite{lia99}. Related works are those where the covariate measured with error is $\mathcal{X}$ instead of $X$; for instance, \cite{lia00} dealt with such a case and showed that the resulting kernel-based estimator of $\pmb{\beta}_0$ is asymptotically normal. 

Focusing now on the SFPLME model (functional $\mathcal{X}$), the only paper we know is that of \cite{zhu20}. They extended to the functional setting the estimators proposed by \cite{lia99}, and established the asymptotic normality and the rate of uniform convergence of the corresponding estimators of $\pmb{\beta}_0$ and $m(\cdot)$, respectively; in addition, they proposed a test statistics for testing hypothesis on $\pmb{\beta}_0$, and derived its limiting distribution under the null and alternative hypotheses.

Except \cite{novo_2019}, the rest of the papers named in the previous paragraphs obtained results for kernel-based estimators. An important point when dealing with infinite-dimensional data (in particular, with functional data) is the need for taking into account local structures of the data. Nevertheless, the bandwidth (say $h$) used in kernel-based estimators does not depend on the covariate in the nonparametric component ($\chi$); so, it does not capture such local features. Another popular technique used in nonparametric and semiparametric statistics considers $k$ nearest neighbour ($k$NN) ideas. Such ideas are especially recommended when data have local structures, because $k$NN-based estimators involve a local bandwidth (that is, depending on $\chi$) making it possible to capture local features ($k$NN gives location adaptive methods). See, for instance, \cite{kara_JMVA}, \cite{novo_2019} and \cite{linav20} for some recent papers developing asymptotic theory based on $k$NN ideas in functional nonparametric, single-index and partial linear regression models, respectively. To the best of our knowledge, in the statistical literature there are no papers dealing with inference based on $k$NN ideas in the SFPLME model.

This paper aims to obtain some first asymptotic results related to $k$NN estimation in the SFPLME model. More precisely, assuming that (\ref{EM}) holds, we construct $k$NN-based estimators for both the vector parameter $\pmb{\beta}_0$ and the operator $m(\cdot)$ in (\ref{eq_model1}); then, the asymptotic distribution and the law of the iterated logarithm for the estimator of $\pmb{\beta}_0$ are obtained, as well as the rate of uniform convergence of the estimator of $m(\chi)$. 

 At this moment, it is worth highlighting the main similarities and differences between \cite{zhu20} and this paper. Both papers extend  the estimators proposed by \cite{lia99} to the functional setting; however, \cite{zhu20} considered kernel-based estimators (as in \citealt{lia99}) while we propose $k$NN-based ones. Regarding the methodological approach adopted in the proofs of the main theoretical results, both papers follow the same approach. In fact, this approach is the standard one in the literature on partial linear regression when the aim is to extend results from the scalar setting to the functional one. Essentially, some key results are required: on the one hand, the rate of uniform convergence of the estimator of $m(\chi)$ in a pure functional nonparametric model; on the other hand, functional versions of certain lemmas used in proving the results obtained in the scalar setting. 

Of course, both the rate of convergence and the lemmas are tied to the type of estimator considered. In this regard, \cite{zhu20} leveraged the rate of convergence and two lemmas from \cite{ferv04} and \cite{anev06}, respectively, while in this paper we use the rate of convergence and two lemmas in \cite{Kud13} and \cite{linav20}, respectively. Other key lemmas used in the proofs related to kernel-based and $k$NN-based estimators are proven in \cite{zhu20} and this paper, respectively. It is also worth noting that this paper extends \cite{linav20} from the case where the covariate $X$ is measured without error to the case where it is measured with error. As expected, the challenges arising in this extension centre on controlling the effect of such errors; once again, our extension of the lemmas from \cite{lia99} from the scalar setting to the functional one allows us to stablish the impact of the measurement error.

The outline of this paper is as follows. In Section \ref{estim}, we propose the estimators, while the assumptions to obtain our asymptotic results are presented in Section \ref{assum}. Section \ref{asymp} is devoted to the statement of the main results. Section \ref{simul} reports a simulation study and Section \ref{reald} presents an application to real data. Some concluding comments are given in Section \ref{conclu}. All proofs are delayed until Appendix \ref{app}, which also contains some technical lemmas.

\section{The $k$NN-based estimators}\label{estim}

This section is devoted to the construction of $k$NN-based estimators for $\pmb{\beta}_0$ and $m(\chi)$ in the SFPLME model (\ref{eq_model1})-(\ref{EM}). For that, we first introduce some conditions on the samples linked to the estimation procedure, as well as some notation.

We assume that $\{(X_i,\mathcal{X}_i,Y_i)\}_{i=1}^n$ are independent and identically distributed (iid) as $(X,\mathcal{X},Y)$ in model (\ref{eq_model1}); $\{(X_i,U_i,W_i)\}_{i=1}^n$  are identically distributed 
as $(X,U,W)$ in model (\ref{EM}); and the measurement errors $\{U_i\}_{i=1}^n$ are independent and independent of $\{(X_i,\mathcal{X}_i,Y_i)\}_{i=1}^n$. Finally, we will denote $\mathbf {X}=\left(X_1,\dots,X_n\right)^{\top}$, $\mathbf {W}=\left(W_1,\dots,W_n\right)^{\top}$, $\mathbf {U}=\left(U_1,\dots,U_n\right)^{\top}$ and $\mathbf {Y}=\left(Y_1,\dots,Y_n\right)^{\top}.$

Researchers in early nonparametric one-dimensional literature utilised the $k$NN ideas for constructing location-adaptive smoothers (see eg \citealt{collomb} or \citealt{dev94}), and they have recently been extended for nonparametric and semiparametric FDA (see eg \citealt{biacg10}, \citealt{kara_JMVA}, \citealt{novo_2019} and \citealt{linav20} for recent results, and Section 2.2 in \citealt{linv18} for a survey). 
To use such ideas to propose $k$NN-based estimators for the SFPLME model (\ref{eq_model1})-(\ref{EM}), it is interesting to note that
$$m(\mathcal{X})=\mathbb{E}\left(Y-X^{\top}\pmb{\beta}_0|\mathcal{X}\right)=\mathbb{E}\left(Y-W^{\top}\pmb{\beta}_0|\mathcal{X}\right).$$ Therefore, to estimate the nonparametric component $m(\chi)$, one can use the observed values of $W_i$ instead of the unobserved values of $X_i$. Then, $k$NN ideas are used for estimating $m(\chi)$ from a nonzero smoothing factor $k=k_n\in \mathbb{N}$ and a kernel function $K$ as follows:
\begin{equation}
	\widehat{m}_{k,\pmb{\beta}_0}(\chi)=\sum_{i=1}^n\omega_{k}(\chi,\mathcal{X}_i)\left(Y_i-W_i^{\top}\pmb{\beta}_0\right)
	,\label{kNN_est1}
\end{equation}
\noindent where, $\forall \chi\in S_\mathcal{F}$, we have denoted
\begin{equation}
	\omega_{k}(\chi,\mathcal{X}_i)=\frac{K\left(H_{k,\chi}^{-1}d\left(\mathcal{X}_i,\chi\right)\right)}{\sum_{i=1}^nK\left(H_{k,\chi}^{-1}d\left(\mathcal{X}_i,\chi\right)\right)},
	\label{pesos}
\end{equation}
with
$$
H_{k,\chi}=\min\left\{h\in \mathbb{R}^+ \mbox{\text{ such that }} \sum_{i=1}^n1_{B(\chi,h)}(\mathcal{X}_i)=k\right\}, 
$$
where $B(\chi,h)=\left\{z\in \mathcal{F}:d\left(\chi,z\right) \leq h\right\}$. 

Actually, (\ref{kNN_est1}) is an infeasible estimator for $m(\chi)$ because it depends on the unknown parameter vector $\pmb{\beta}_0$. Then, the usual is to obtain an estimator of $\pmb{\beta}_0$ and introduce it in (\ref{kNN_est1}). For that, a first (naive) idea could be to ignore the measurement error (that is, replace $X_i$ by $W_i$ in (\ref{eq_model1})) and then to consider an estimator of $\pmb{\beta}_0$ based on an SFPL model (so, without measurement error); as noted in \cite{carroll95} and \cite{lia99}, among others, such resulting estimator is inconsistent for $\pmb{\beta}_0$. 

A consistent estimator for ${\pmb{\beta}_0}$ could be constructed from the correction for attenuation method proposed by \cite{carroll95} in the context of linear models (see also \citealt{lia99} for the case of non-functional partial linear models). In our case of SFPLME models to be estimated using $k$NN ideas, a natural extension of their proposals is to estimate ${\pmb{\beta}_0}$ through the vector $\widehat{\pmb{\beta}}_{0k}$ that minimizes the score function
\begin{equation}
	\mathcal{Q}_k\left(\pmb{\beta}\right)=\frac{1}{n}\sum_{i=1}^n\left(Y_i - W_i^{\top}\pmb{\beta}- \widehat{m}_{k,\pmb{\beta}}(\mathcal{X}_i)\right)^2 -  \pmb{\beta}^{\top} \pmb{\Sigma}_{uu} \pmb{\beta}. 
	\label{func_minimizar}
\end{equation}
The correction for attenuation aims to reduce the impact of measurement errors that weaken the relationship between the covariates and the response, resulting in smaller coefficients. The purpose of this correction is opposite to the penalization in shrinkage methods for regression. That is why the negative symbol appears in the second term. Note that, taking into account that the function $\mathcal{Q}_k(\cdot)$ can be written as
\begin{equation}
	\mathcal{Q}_k\left(\pmb{\beta}\right)=\frac{1}{n}\left(\widetilde{\mathbf {Y}}-\widetilde{\mathbf {W}}\pmb{\beta}\right)^{\top}\left(\widetilde{\mathbf {Y}}-\widetilde{\mathbf {W}}\pmb{\beta}\right)-  \pmb{\beta}^{\top} \pmb{\Sigma}_{uu} \pmb{\beta},
	\label{func_minimizar_2}
\end{equation}
we have that
\begin{equation}
	\widehat{\pmb{\beta}}_{0k}=\left(\widetilde{\mathbf {W}}^{\top}\widetilde{\mathbf {W}}-n\pmb{\Sigma}_{uu} \right)^{-1}\widetilde{\mathbf {W}}^{\top}\widetilde{\mathbf {Y}},
	\label{beta-est}
\end{equation}
where 
for any $(n\times q)$-matrix $\mathbf {A}$ $(q\geq 1)$ and number of neighbours $k$, we have denoted
$\widetilde{\mathbf {A}}=\left(\mathbf{I}-\mathbf {V}_{k}\right)\mathbf {A}, \mbox{ where } \mathbf {V}_{k}=\left(\omega_{k}(\mathcal{X}_i,\mathcal{X}_j)\right)_{i,j}$.

Finally, considering $\widehat{\pmb{\beta}}_{0k}$ in (\ref{kNN_est1}) instead of $\pmb{\beta}_0$, we obtain the feasible estimator of $m$

\begin{equation}
	\widehat{m}_{k}(\chi):=\widehat{m}_{k,\widehat{\pmb{\beta}}_{0k}}(\chi)=\sum_{i=1}^n\omega_{k}(\chi,\mathcal{X}_i)\left(Y_i-W_i^{\top}\widehat{\pmb{\beta}}_{0k}\right)
	.\label{kNN_est}
\end{equation}

\begin{remark}
	It is worth being noted that the $k$NN-based estimators (\ref{beta-est}) and (\ref{kNN_est}) are extensions of the kernel-based ones considered in \cite{zhu20}, where the weights
	\begin{equation}
		\omega_{h}(\chi,\mathcal{X}_i)=\frac{K\left(h^{-1}d(\mathcal{X}_i,\chi)\right)}{\sum_{i=1}^nK\left(h^{-1}d(\mathcal{X}_i,\chi)\right)} \label{pesos-kernel}
	\end{equation}
	were used instead of the weights in (\ref{pesos}) ($h\in \mathbb{R}^+$ is a smoothing parameter known as the bandwidth). The $k$NN-based estimators present, at least, two main advantages in practice in comparison with  the kernel ones. On the one hand,
	although the number of neighbours, $k$, is fixed, the bandwidth $H_{k,\chi}$ varies with $\chi$, providing the local-adaptive property of $k$NN-based estimators (allowing adaptation to heterogeneous designs). On the other hand, the selection of the smoothing parameter $k$ has a lower computational cost than the selection of $h$, since $k$ takes values in the finite set $\{1,2,\dots,n\}$. However, the price to pay for these nice practical features is that, from a theoretical point of view, properties of the $k$NN statistics are much more difficult to obtain, mainly because  $H_{k,\chi}$ is a random variable depending on $\mathcal{X}_i$ ($i=1,\dots,n$) and avoiding for decomposing (\ref{kNN_est}) as sums of iid terms.
\end{remark}

\section{Technical assumptions}\label{assum}

First of all, let us denote $X_i=(X_{i1}, \ldots,X_{ip})^{\top}$, $g_j(\mathcal{X}_{i})=\mathbb{E}(X_{ij}|\mathcal{X}_{i})$, $\eta_{ij}=X_{ij}-g_j(\mathcal{X}_{i})$, $\eta_i=(\eta_{i1}, \ldots, \eta_{ip})^{\top}$ and $U_i=(U_{i1}, \ldots,U_{ip})^{\top}$, ($i=1,\ldots,n; j=1,\ldots,p$). In addition, for any set $S \subset \mathcal{F}$ and $\epsilon>0$, $\psi_{S}(\epsilon)$ denotes the Kolmogorov's $\epsilon$-entropy of $S$, which is defined as $\psi_{S}(\epsilon)=\log (N_{\epsilon}(S))$, where $N_{\varepsilon}(S)$ is the minimal number of open balls in $\mathcal{F}$ of radius $\epsilon$ which is necessary to cover $S$.

In order to prove our asymptotic results, we will use the following assumptions:

\noindent (A1) $\forall \epsilon >0,$ $\varphi_{\chi}(\epsilon):=P(\mathcal{X}\in B(\chi,\epsilon))>0$, with $\varphi_{\chi}(\cdot)$ continuous on a neighbourhood of $0$ and $\varphi_{\chi}(0)=0.$

\noindent (A2) There exist a nonnegative function $\phi(\cdot)$ regularly varying at $0$ with nonnegative index, a positive function $g(\cdot)$ and a positive number $\alpha$ such that:
\begin{itemize}
	\item[(i)]$\phi (0)=0$ and $\lim_{\epsilon\rightarrow 0}\phi (\epsilon)=0.$
	\item[(ii)]$\exists ~C>0$ and $\exists \ \eta_{0}>0$ such that, $\forall 0<\eta<\eta_{0},$ $\phi^{'}(\eta)<C.$
	\item[(iii)]$\sup_{\chi\in S_{\mathcal{F}}}|\frac{\varphi_{\chi}(\epsilon)}{\phi (\epsilon)}-g(\chi)|=O(\epsilon^{\alpha}) \ \text{as} \ \epsilon \rightarrow 0$.
	\item[(iv)]$\exists C<\infty $ such that
	$\forall \text{ }\left( u,v\right) \in S_{\mathcal{F}}\times S_{\mathcal{F}},\ \forall \text{ }%
	f\in \left\{ m,g_{1},\ldots,g_{p}\right\} ,\text{ }\left\vert f\left( u\right)
	-f\left( v\right) \right\vert \leq Cd\left( u,v\right) ^{\alpha }.$
\end{itemize}
(A3) The kernel function, $K(\cdot)$, satisfies:
\begin{itemize}
	\item[(i)]$K(\cdot)$ is a nonnegative, bounded and non increasing function with support $[0,1]$ and Lipschitz on $[0,1).$
	\item[(ii)]If $K(1)=0$, it must also be such that $-\infty<C<K'(t)<C'<0.$
\end{itemize}
(A4) The Kolmogorov's $\epsilon$-entropy of $S_{\mathcal{F}}$ satisfies:
\begin{equation}
	\sum_{n=1}^{n}\exp\{(1-\omega)\psi_{S_{\mathcal{F}}}(\frac{\log n}{n})\}<\infty \text{ for some } \omega >1. \nonumber
\end{equation}
(A5) $k=k_{n}$ is a sequence of positive real numbers such that:
\begin{itemize}
	\item[(i)]$\frac{k}{n}\rightarrow 0$ and $ \frac{\log n}{k}\rightarrow 0$ as $n\rightarrow \infty$.
	\item[(ii)]For $n$ large enough, $\frac{(\log n)^{2}}{k}<\psi_{S_{\mathcal{F}}}(\frac{\log n}{n})<\frac{k}{\log n}.$
\end{itemize}
(A6) Moment conditions:
\begin{itemize}
	\item[(i)]
	$\forall r\geq 3, \ 1\leq j\leq p$ and $\chi \in S_{\mathcal{F}},$ 
	$\mathbb{E}(|Y_{1}|^{r}|\mathcal{X}_{1}=\chi)\leq C_{1}<\infty$, $\mathbb{E}(|X_{1j}|^{r}|\mathcal{X}_{1}=\chi  )\leq C_{2}<\infty$ and $\mathbb{E}(|U_{1j}|^{r})\leq C_{3}<\infty.$
	\item[(ii)]%
	$\mathbf {B}=\mathbb{E}\left(\eta_{1}\eta_{1}^{\top}\right)$ is a positive definite matrix. 
\end{itemize}
\begin{remark}
	As can be seen in (\ref{beta-est}) and (\ref{kNN_est}), nonparametric smoothing plays a main role in the proposed estimators $\widehat{\pmb{\beta}}_{0k}$ and $\widehat{m}_{k}(\chi)$. In fact, rates of uniform convergence of nonparametric estimators are repeatedly used to get our asymptotic results (see Appendix \ref{app}). The aim of assumptions (A1)-(A6)(i) is to achieve such rates. This kind of assumptions was used in a context of functional nonparametric regression models in, for instance, \cite{ferltv10} (kernel-based estimation) and \cite{Kud13} ($k$NN-based estimation); for comments on these hypotheses, see \cite{Kud13}. Assumption (A6)(ii) is usually impossed to ensure identifiability in the SFPL model (\ref{eq_model1}); see, for instance, \cite{anev06}. Note that, excepting the moment condition on the measurement error (last part in Assumption A6(i)), all the other conditions were used in \cite{linav20} and certainly they are not restrictive.
\end{remark}

\section{Asymptotic results}\label{asymp}

The next Theorem \ref{theorem1} is the main contribution of this paper.  

\begin{theorem} \label{theorem1} Under assumptions (A1)-(A6), if in addition $\frac{\sqrt{n}\log^{2}n}{k}\rightarrow 0$, $\sqrt{n}\phi^{-1}(\frac{k}{n})^{\alpha}\rightarrow 0$, $\frac{\sqrt{n}\psi_{S_{\mathcal{F}}}(\frac{\log n}{n})}{k}\rightarrow 0$ as $n\rightarrow \infty$ and $k\geq n^{(2/r)+b}/\log ^{2}n$ for $n$ large enough and some constant $b>0$ with $\frac{2}{r}+b>\frac{1}{2}$ (where $r\geq 3$), then we have:
	
	\begin{itemize}
		\item[(i)]$\sqrt{n}(\widehat{\pmb{\beta}}_{0k}-\pmb{\beta}_0)\rightarrow N(0,\mathbf {B}^{-1}  \pmb{\varGamma}  \mathbf {B}^{-1})$, 
		
		where
		$$
		\pmb{\varGamma}=\mathbb{E}\left(\varepsilon-\mathbf {U}^{\top}\pmb{\beta}_0\right)^2\mathbf {B}+\mathbb{E}\left\{\left(\mathbf {U}^{\top}\mathbf {U}-\pmb{\Sigma}_{uu}\right)\pmb{\beta}_0\right\}^{\otimes 2}+\pmb{\Sigma}_{uu} \sigma^2,
		$$
		with $\mathbf {M}^{\otimes 2}=\mathbf {M}\mathbf {M}^{\top}$ for any matrix $\mathbf {M}.$
		\item[(ii)]$\limsup_{n\rightarrow \infty}\big(\frac{n}{2\log \log n}\big)^{\frac{1}{2}}|\widehat{\beta}_{0kj}-\beta_{0j}|=(\sigma_{jj})^{\frac{1}{2}} \ a.s. \ (j=1,\ldots,p)$,
		
		where $\beta_{0j}$ and $\widehat{\beta}_{0kj}$ denote the $j$-th components of the vectors $\pmb{\beta}_0$ and $\widehat{\pmb{\beta}}_{0k}$, respectively, while $\sigma_{jj}$ denotes the $j$-th element of the diagonal of the matrix $\mathbf {B}^{-1}  \pmb{\varGamma}  \mathbf {B}^{-1}.$
		\item[(iii)]$\sup_{\chi\in S_{\mathcal{F}}}|\widehat{m}_{k}(\chi)-m(\chi)|=O\left(\phi^{-1}(\frac{k}{n}) ^{\alpha}+\sqrt{\frac{\psi_{S_{\mathcal{F}}}(\frac{\log n}{n})}{k}}\right) \ a.s.$
	\end{itemize}
\end{theorem}

\begin{remark}
	On the one hand, the asymptotic results in Theorem \ref{theorem1} can be seen as an extension of those in \cite{linav20} from an SFPL model (\ref{eq_model1}) to an SFPLME one (\ref{eq_model1})-(\ref{EM}). A nice (and expected) conclusion is that the presence of measurement errors influences the covariance matrix of the limit distribution of the estimator of $\pmb{\beta}_0$, but not the rate of convergence neither of such estimator nor the estimator of $m(\chi)$. On the other hand, our results (i) and (iii) extends Theorems 2.3 and 2.4, respectively, in \cite{zhu20} from kernel-based estimators to $k$NN-based ones. From an asymptotic point of view, both estimators of $\pmb{\beta}_0$ (kernel- or $k$NN-based) have the same covariance matrix; focusing now on the rates of convergence of each estimator of $m(\chi)$, one notes that they depend on both the smoothing parameter ($h$ or $k$) and the topology of ($\mathcal{F}, d(\cdot,\cdot)$). Assumptions on the topology used in this paper are more general than the ones in \cite{zhu20}. If, for the sake of comparison, we consider both the same topology as that in \cite{zhu20} and suitable values for the smoothing parameters, the conclusion is that both estimators converge at the same rate (as expected). 
\end{remark}

\begin{remark}
	\label{unknown}
	It is worth being noted that, in practice, not always the covariance matrix of the measurement error, $\pmb{\Sigma}_{uu}$, is known. In that case, our estimators $\widehat{\pmb{\beta}}_{0k}$ (\ref{beta-est}) and $\widehat{m}_{k}(\chi)$ (\ref{kNN_est}) are infeasible because they depend on $\pmb{\Sigma}_{uu}$; therefore, one needs a consistent estimator for $\pmb{\Sigma}_{uu}$. \cite{carroll95} (Chapter 3) proposed a consistent and unbiased estimator for $\pmb{\Sigma}_{uu}$, which was used, in different settings, in \cite{lia99} (Section 5) and \cite{zhu19} (Remark 4). More specifically, assuming that one has $r>1$ replicates of $W_{i}$ (for notational convenience, we consider here that $r$ does not depend on $i$) measuring the same $X_{i}$ ($i=1,\ldots,n$), $$W_{i}^{(j)}=X_{i}+U_{i}^{(j)} \ (j=1,\ldots,r),$$ the proposed estimator of $\pmb{\Sigma}_{uu}$ was: 
\begin{equation}
\widehat{\pmb{\Sigma}}_{uu}=\frac{1}{n(r-1)}\sum_{i=1}^n\sum_{j=1}^r(W_{i}^{(j)}-\overline{W}_{i}^{(\cdot)})(W_{i}^{(j)}-\overline{W}_{i}^{(\cdot)})^{\top}, \label{Sigma_est}
\end{equation}
where $$\overline{W}_{i}^{(\cdot)}=\frac{1}{r}\sum_{j=1}^rW_{i}^{(j)}.$$ Note that this estimator $\widehat{\pmb{\Sigma}}_{uu}$ can also be used in our setting of SFPLME model (\ref{eq_model1})-(\ref{EM}) (the functional feature of the SFPLME does not influence either the construction or the asymptotic properties of that estimator). Then, feasible estimators of $\pmb{\beta}_{0}$ and $m(\chi)$ are
\begin{equation}
	\widehat{\pmb{\beta}}_{0k}^{\ast}=\left(\widetilde{\mathbf {\overline{W}}}^{\top}\widetilde{\mathbf {\overline{W}}}-\frac{n}{r}\widehat{\pmb{\Sigma}}_{uu} \right)^{-1}\widetilde{\mathbf {\overline{W}}}^{\top}\widetilde{\mathbf {Y}}
	\label{beta-est-replic}
\end{equation}
and
\begin{equation}
	\widehat{m}_{k}(\chi)^{\ast}=\sum_{i=1}^n\omega_{k}(\chi,\mathcal{X}_i)\left(Y_i-\overline{W}_{i}^{(\cdot)\top}\widehat{\pmb{\beta}}_{0k}^{\ast}\right),\label{kNN_est}
\end{equation}
respectively, where $\pmb{\overline{W}}^{(\cdot)}=(\overline{W}_{1}^{(\cdot)},\ldots,\overline{W}_{n}^{(\cdot)})^{\top}$.

Now, focusing  on the impact of the estimation of $\pmb{\Sigma}_{uu}$ in the results of Theorem \ref{theorem1}, and using similar techniques as those applied to prove that theorem, one can show that
$$\sqrt{n}(\widehat{\pmb{\beta}}_{0k}^{\ast}-\pmb{\beta}_0)\rightarrow N(0,\mathbf {B}^{-1}  \pmb{\varGamma}^{\ast}  \mathbf {B}^{-1}),$$ where
$$\pmb{\varGamma}^{\ast}=\mathbb{E}\left(\varepsilon-{\overline{U}}^{(\cdot)\top}\pmb{\beta}_0\right)^2\mathbf {B}+\mathbb{E}\left\{\left({\overline{U}}^{(\cdot)\top}{\overline{U}}^{(\cdot)}-\frac{1}{r} \pmb{\Sigma}_{uu}\right)\pmb{\beta}_0\right\}^{\otimes 2}+ \sigma^2\mathbb{E}\left( {\overline{U}}^{(\cdot)\top} {\overline{U}}^{(\cdot)}\right)$$ (in a similar way as $\overline{W}_{i}^{(\cdot)}$, ${\overline{U}}^{(\cdot)}$ denotes the mean of $r$ $U$'s). In addition, both the law of the iterated logarithm for $\widehat{\pmb{\beta}}_{0k}^{\ast}$ (considering $(\sigma^{\ast}_{ij})=\mathbf {B}^{-1}\pmb{\varGamma}^{\ast}\mathbf {B}^{-1}$ instead of $(\sigma_{ij})=\mathbf {B}^{-1}\pmb{\varGamma}\mathbf {B}^{-1}$ in Theorem \ref{theorem1}(ii)) and the rate of uniform convergence of $\widehat{m}_{k}(\cdot)^{\ast}$ remain valid.
\end{remark}

\section{Simulation study}\label{simul}

The aim of this section is to illustrate the effectiveness of the proposed estimator (\ref{beta-est}) in reducing the estimation bias when finite sample sizes are used. Specifically, we compared the accurate of $\widehat{\pmb{\beta}}_{0k}$ with that of different estimators of $\pmb{\beta}_0$ in the SFPLME model (\ref{eq_model1})-(\ref{EM}), as well as the predictive power of the corresponding estimated SFPLME models. Both $k$NN- and kernel-based versions of all the considered estimators were included in the comparative study. 

\subsection{The design}
For different values of $n$, iid observations $\{(X_i,\mathcal{X}_i,Y_i)\}_{i=1}^{n+100}$ were generated from the SFPL model
\begin{equation*}
	Y=X^{\top}\pmb{\beta}_0+m\left(\mathcal{X}\right)+\varepsilon ,
\end{equation*}
where $X_i=(X_{i1},X_{i2})$ with $X_{i1}$ and $X_{i2}$ being $N(1,1)$. The functional covariate was
\begin{equation*}
	\mathcal{X}_i(t)=a_i(t-0.5)^2+b_i \ (t \in [0,1]),
\end{equation*}
where we considered a mixture distribution for the random variable $a_i$: $U(-3,3)$ with probability 0.5, and $U(20,21)$ with probability 0.5; the distribution of the random variable $b_i$ was $N(0,1)$. Each curve $\mathcal{X}_i$ was discretized in $100$ equispaced points ($0=t_1<t_2< \cdots <t_{100}=1$). The considered vector parameter $\pmb{\beta}_0$ was $(-1,1.5)^{\top}$, while the operator $m(\cdot)$ was 
\begin{equation*}
	m(\mathcal{X})=\int_0^1(\mathcal{X}^{(1)}(t))^2dt.
\end{equation*}
In addition, the distribution of the random error, $\varepsilon_i$, was $N(0,0.5^2)$. Finally, iid observations $\{U_i\}_{i=1}^{n}$ were generated from a distribution $N(0,\pmb{\Sigma}_{uu})$ (recall that we observe $W_i=X_i+U_i$ instead of $X_i$). Note that the design of this simulation study is similar to that in \cite{zhu20}, except that in \cite{zhu20} the considered distribution  for the random variable $a_i$ was $U(-3,3)$ instead of a mixture one. 
The role of such a mixture distribution was to introduce heterogeneity in the sample of the functional covariate $\mathcal{X}$. 

\subsection{The study}

To assess the finite-sample performance of the proposed Correction-for-Attenuation-based estimator (\ref{beta-est}), $\widehat{\pmb{\beta}}_{0k}$  (CfA), we compared it with that of the naive estimator, $\breve{\pmb{\beta}}_{0k}$ (NAIVE), which ignores measurement error, and the oracle estimator, $\widetilde{\pmb{\beta}}_{0k}$ (ORACLE), which assumes that $X$ can be observed exactly (note that the oracle estimator was proposed and studied in \citealt{linav20}). Specifically, the expressions of such estimators are:
\begin{equation}
	\breve{\pmb{\beta}}_{0k}=\left(\widetilde{\mathbf {W}}^{\top}\widetilde{\mathbf {W}}\right)^{-1}\widetilde{\mathbf {W}}^{\top}\widetilde{\mathbf {Y}}
	\label{beta-naive}
\end{equation}
and
\begin{equation}
	\widetilde{\pmb{\beta}}_{0k}=\left(\widetilde{\mathbf {X}}^{\top}\widetilde{\mathbf {X}}\right)^{-1}\widetilde{\mathbf {X}}^{\top}\widetilde{\mathbf {Y}}.
	\label{beta-oracle}
\end{equation}
Note that the three estimators, $\widehat{{\pmb{\beta}}}_{0k}$, $\breve{\pmb{\beta}}_{0k}$ and $\widetilde{\pmb{\beta}}_{0k}$, are $k$NN-based estimators. Kernel-based versions of $\widehat{{\pmb{\beta}}}_{0k}$, $\breve{\pmb{\beta}}_{0k}$ and $\widetilde{\pmb{\beta}}_{0k}$ were also considered in this study. They were denoted as $\widehat{{\pmb{\beta}}}_{0h}$, $\breve{\pmb{\beta}}_{0h}$ and $\widetilde{\pmb{\beta}}_{0h}$, respectively. Their construction differs from that based on $k$NN only in the use of weigths (\ref{pesos-kernel}) instead of (\ref{pesos}). 

Note that such weights were used to define $\widetilde{\mathbf {A}}$, where $\mathbf {A}=\mathbf {X},\mathbf {W}$ or $\mathbf {Y}$. 
We employed the asymetrical quadratic kernel $K(u)=1.5(1-u^2)1_{[0,1]}(u)$. Given the smoothness of curves $\mathcal{X}_i$, we considered a semi-metric based on derivatives (see Section 13.6 in \citealt{ferv06}); specifically, we used the class of semi-metrics:
\begin{equation}
	d_q(\mathcal{X}_i,\mathcal{X}_j)=\left(\int_0^1(\mathcal{X}_i^{(q)}(t)-\mathcal{X}_j^{(q)}(t))^2dt\right)^{1/2}, \ q=0,1,2.
	\label{X-sim}
\end{equation}

In the first stage, we utilized the training sample ${\cal{S}}_{n,train}=\{(W_i,\mathcal{X}_i,Y_i)\}_{i=1}^{n}$ (${\cal{S}}_{n,train}=\{(X_i,\mathcal{X}_i,Y_i)\}_{i=1}^{n}$) to obtain the estimates $\widehat{{\pmb{\beta}}}_{0k}$, $\breve{\pmb{\beta}}_{0k}$, $\widehat{{\pmb{\beta}}}_{0h}$ and $\breve{\pmb{\beta}}_{0h}$ ($\widetilde{\pmb{\beta}}_{0k}$ and $\widetilde{\pmb{\beta}}_{0h}$). We derived the estimates $\widehat{m}(\cdot)$ and $\breve{m}(\cdot)$ ($\widetilde{m}(\cdot)$) associated to each estimate of $\pmb{\beta}_{0}$ too. We also used the training sample to select the tuning parameters: $k$, $h$ and $q$, using cross-validation. 

Then, in the second stage, we used the testing sample ${\cal{S}}_{n,test}=\{(X_j,\mathcal{X}_j,Y_j)\}_{j=n+1}^{n+100}$ to measure the quality of the predictions of the response variable through the Mean Square Error of Prediction (MSEP):
\begin{equation}
	MSEP_n=\frac{1}{100}\sum_{j=n+1}^{n+100}\left(Y_j - X^{\top}_j\overline{\pmb{\beta}}_{0}-\overline{m}_{\overline{\pmb{\beta}}_{0}}(\mathcal{X}_j)\right)^2,
	\label{MSEP-sim}
\end{equation}
where $\overline{\pmb{\beta}}_{0}$ and $\overline{m}_{\overline{\pmb{\beta}}_{0}}(\cdot)$ denote each pair of estimates of $\pmb{\beta}_{0}$ and $m(\cdot)$ obtained in the fist stage. Note that (\ref{MSEP-sim}) was defined in a similar way to those in \cite{zhu19} (page 306).

Four sample sizes, $n$, were considered: $100$, $200$, $400$ and $800$. Regarding the covariance matrices of the measurement errors, $\pmb{\Sigma}_{uu}$, we examined three low-moderate and three moderate-high measurement error scenarios. The first scenario was composed of the matrices diag$(0.1^2,0.1^2)$, diag$(0.15^2,0.15^2)$ and diag$(0.2^2,0.2^2)$, while the matrices in the second scenario were diag$(0.1,0.1)$, diag$(0.2,0.2)$ and diag$(0.4,0.4)$ (recall that the covariance matrix of the covariate measured without error, $X_i$, was diag$(1,1)$). For each of the 24 considered combinations $(n,\pmb{\Sigma}_{uu})$, we replicated the experiment $M=100$ times.

\subsection{The results}
\label{results-sim}
\begin{figure}[]
	\centering

	\includegraphics[width=0.45\textwidth]{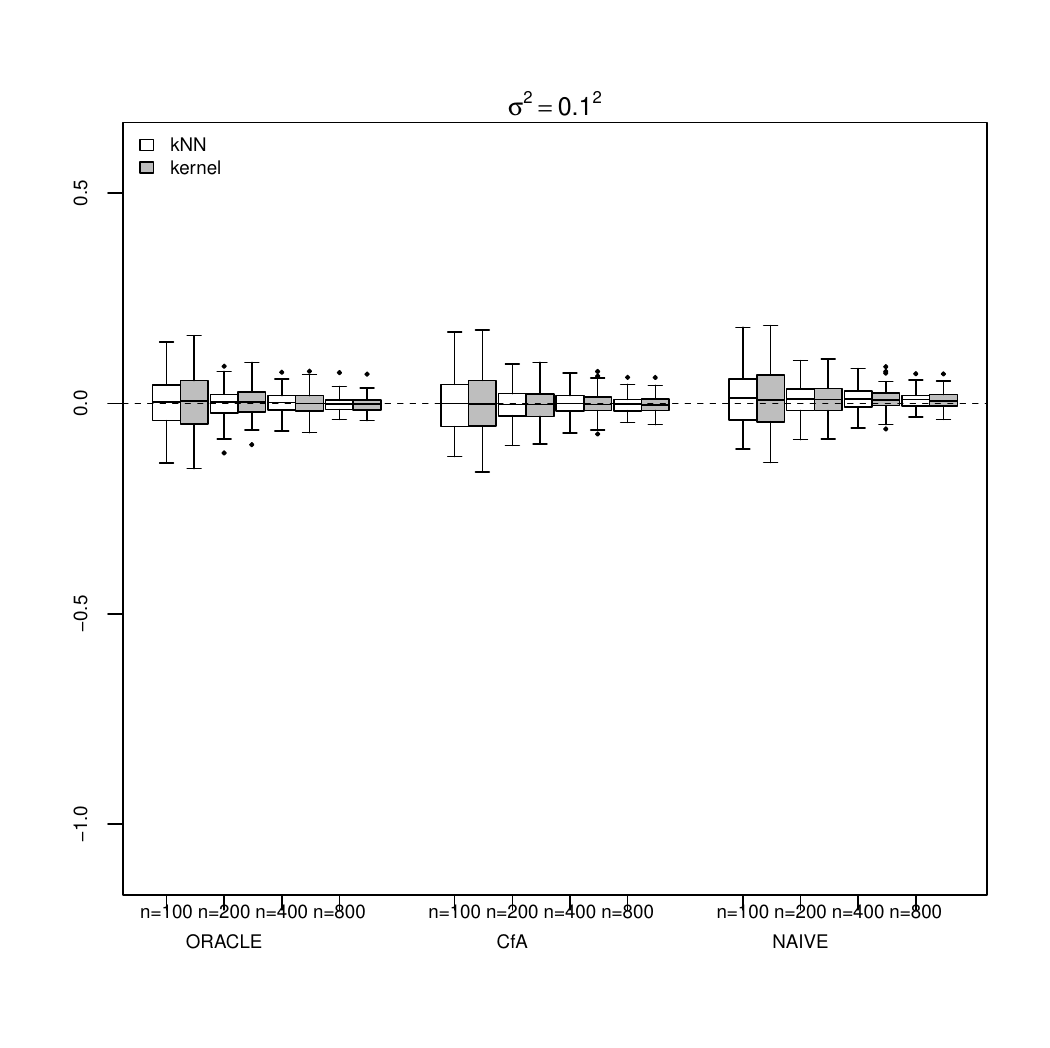} %
	\includegraphics[width=0.45\textwidth]{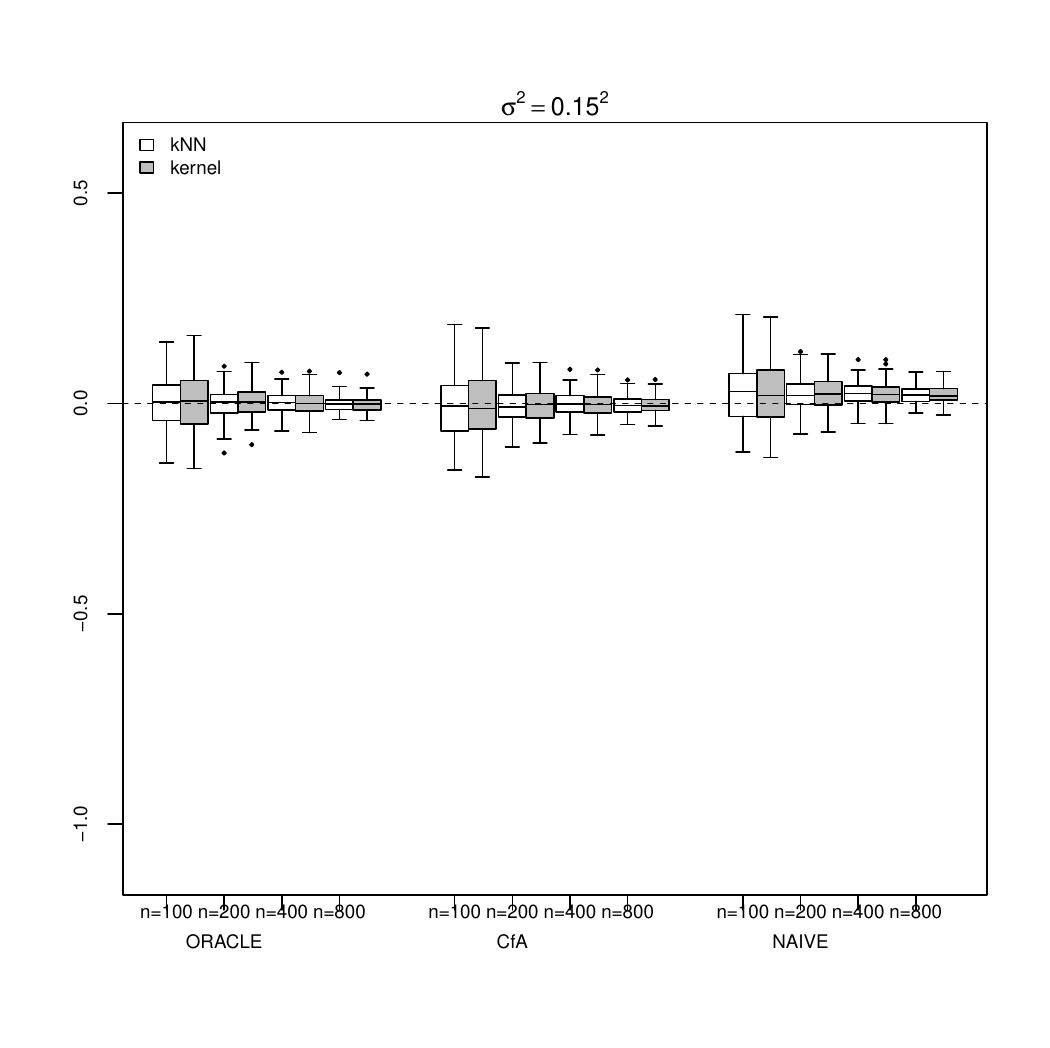}\newline
	\includegraphics[width=0.45\textwidth]{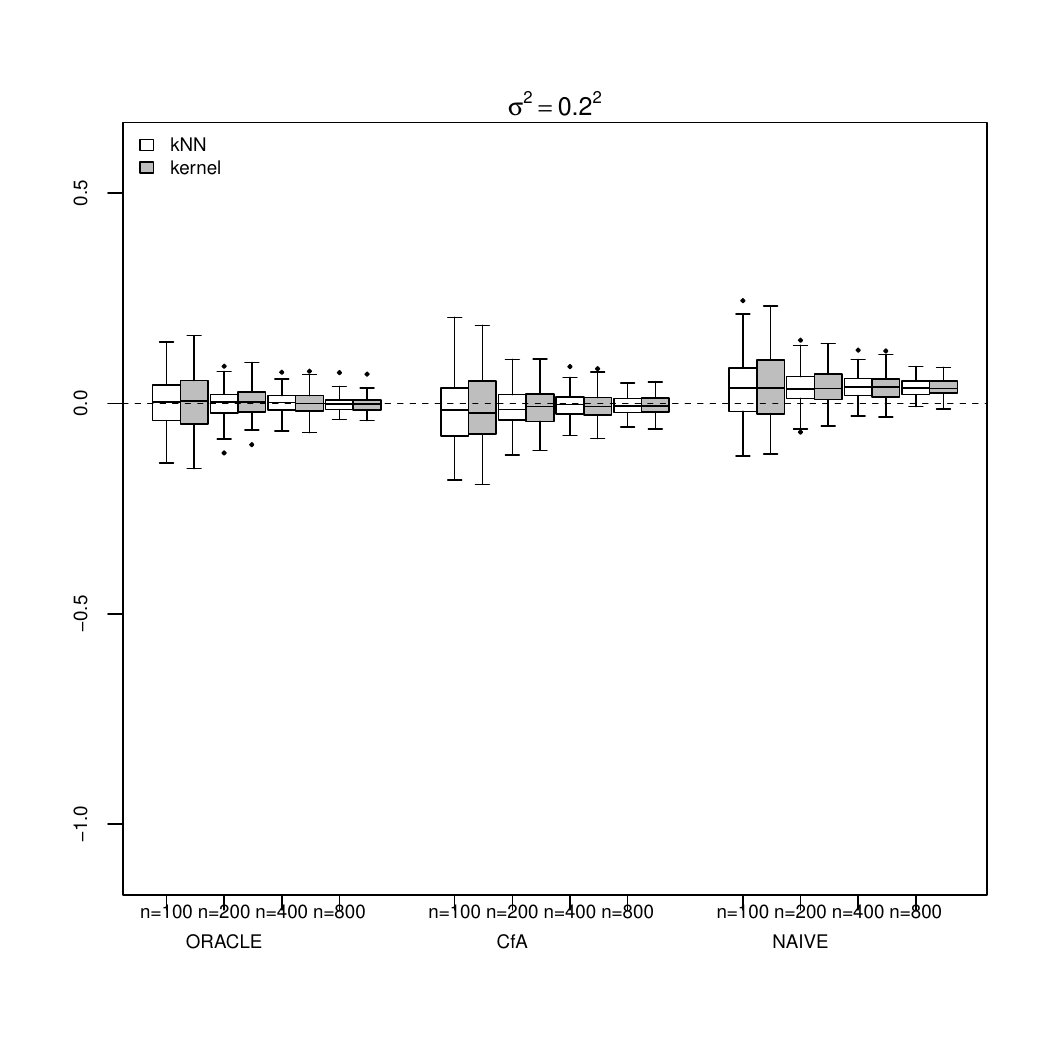} %
	\includegraphics[width=0.45\textwidth]{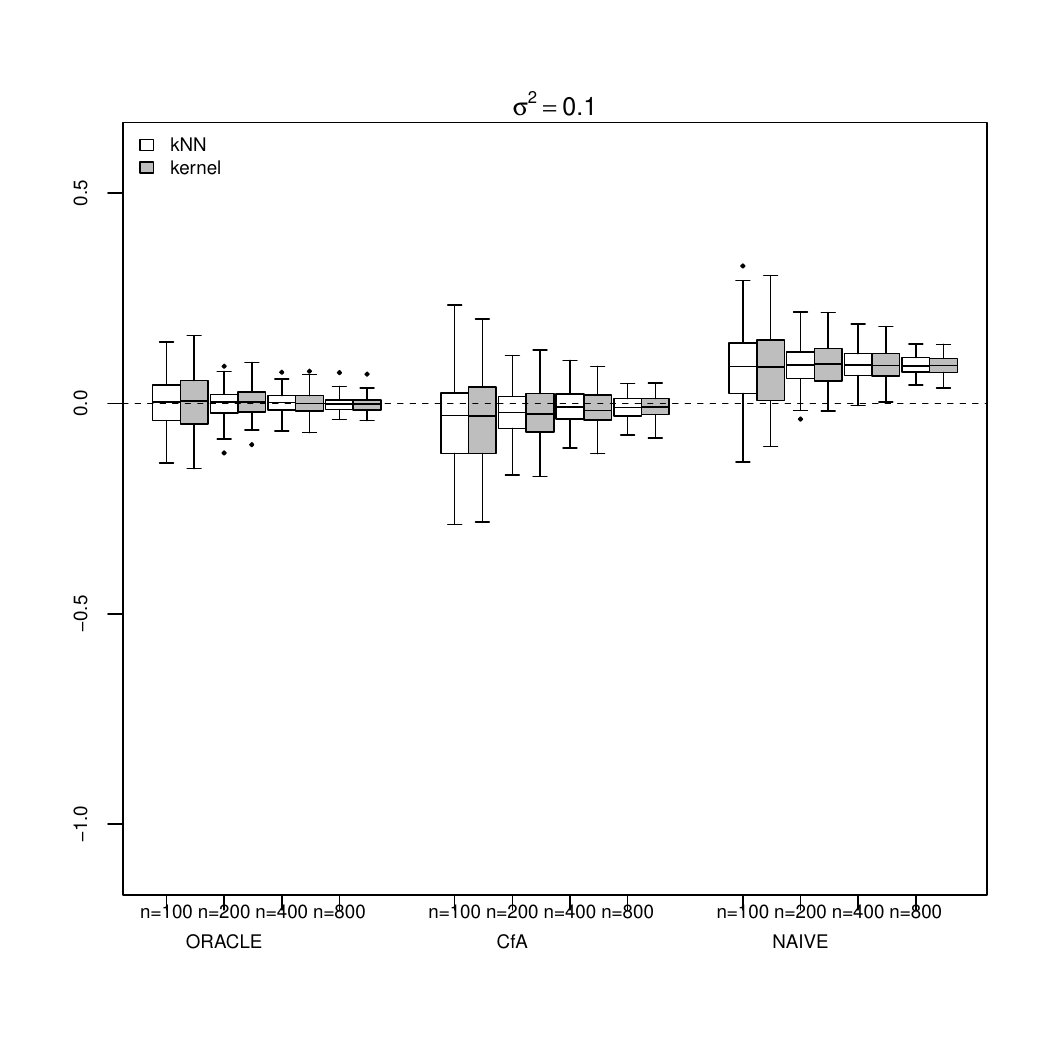}\newline
	\includegraphics[width=0.45\textwidth]{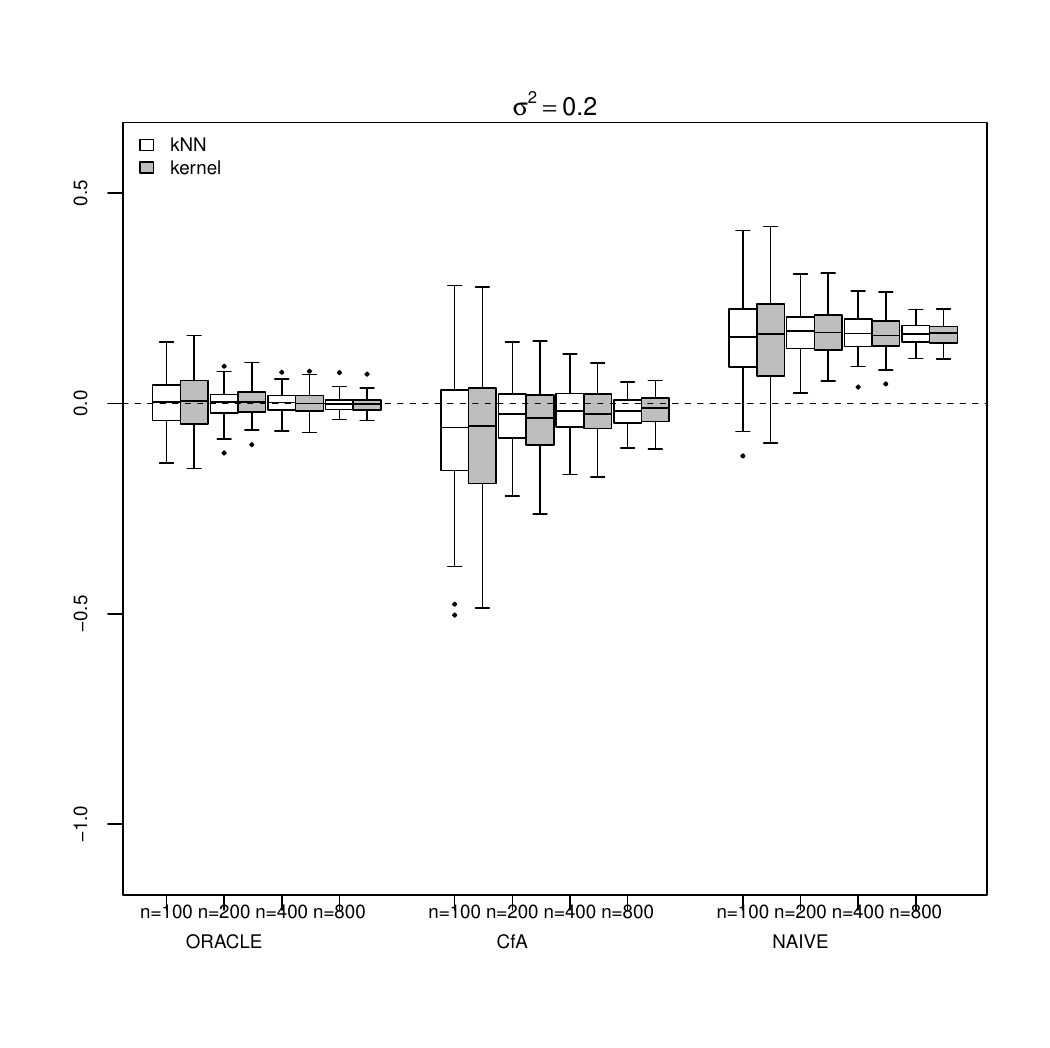} %
	\includegraphics[width=0.45\textwidth]{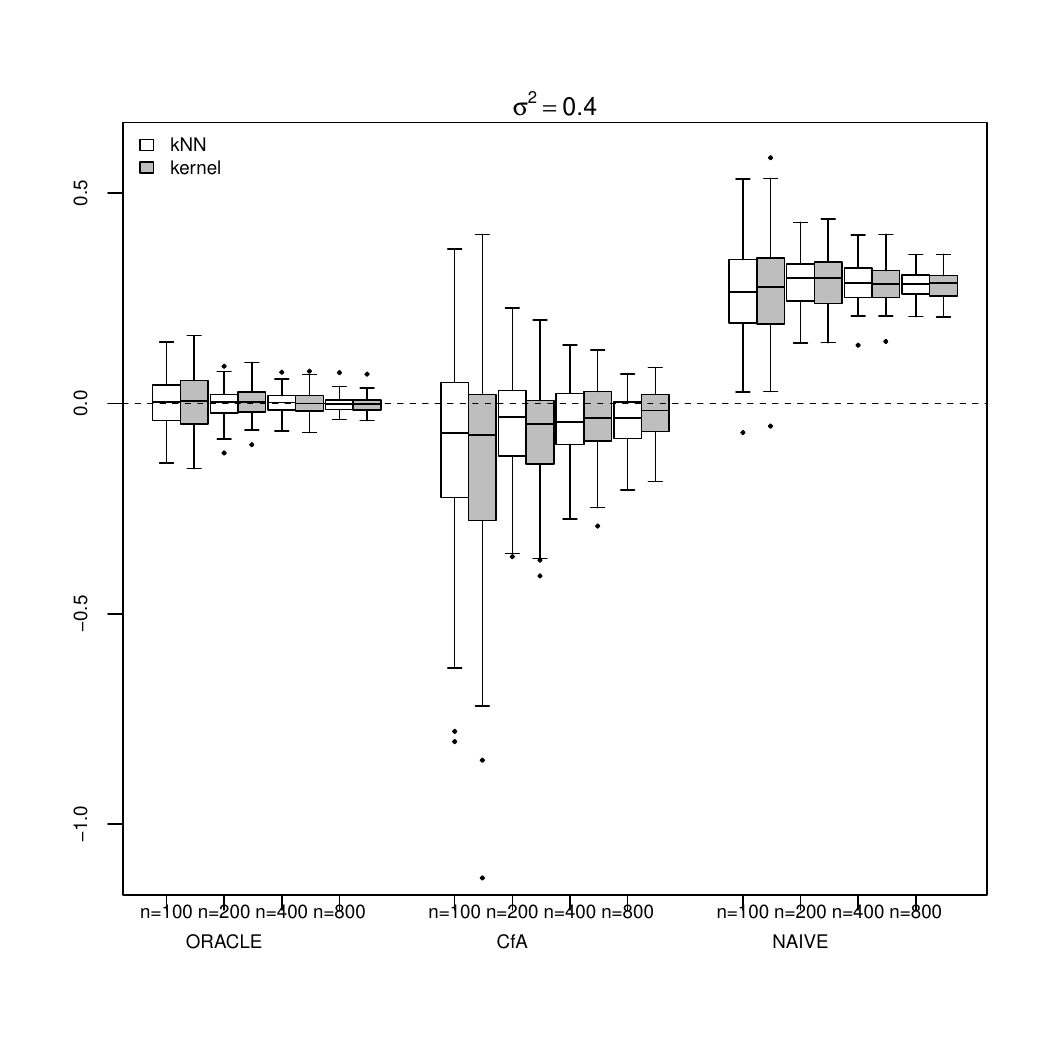}\newline
	
	\vspace{-1cm}
	\caption{Boxplots of $\overline{{\beta}}_{01}-\beta_{01}$ when ORACLE ($\overline{\pmb{\beta}}_{0}=\widetilde{\pmb{\beta}}_{0}$), CfA ($\overline{\pmb{\beta}}_{0}=\widehat{\pmb{\beta}}_{0}$) and NAIVE ($\overline{\pmb{\beta}}_{0}=\breve{\pmb{\beta}}_{0}$) procedures were used. For each of these three classes, both $k$NN- and kernel-based estimators were considered. Different sample sizes, $n$, and covariance matrices of the measurement error, $\pmb{\Sigma}_{uu}=\text{diag}(\sigma^2,\sigma^2)$, were used.}
	\label{fig1}
\end{figure}

\begin{figure}[]
	\centering
	\includegraphics[width=0.45\textwidth]{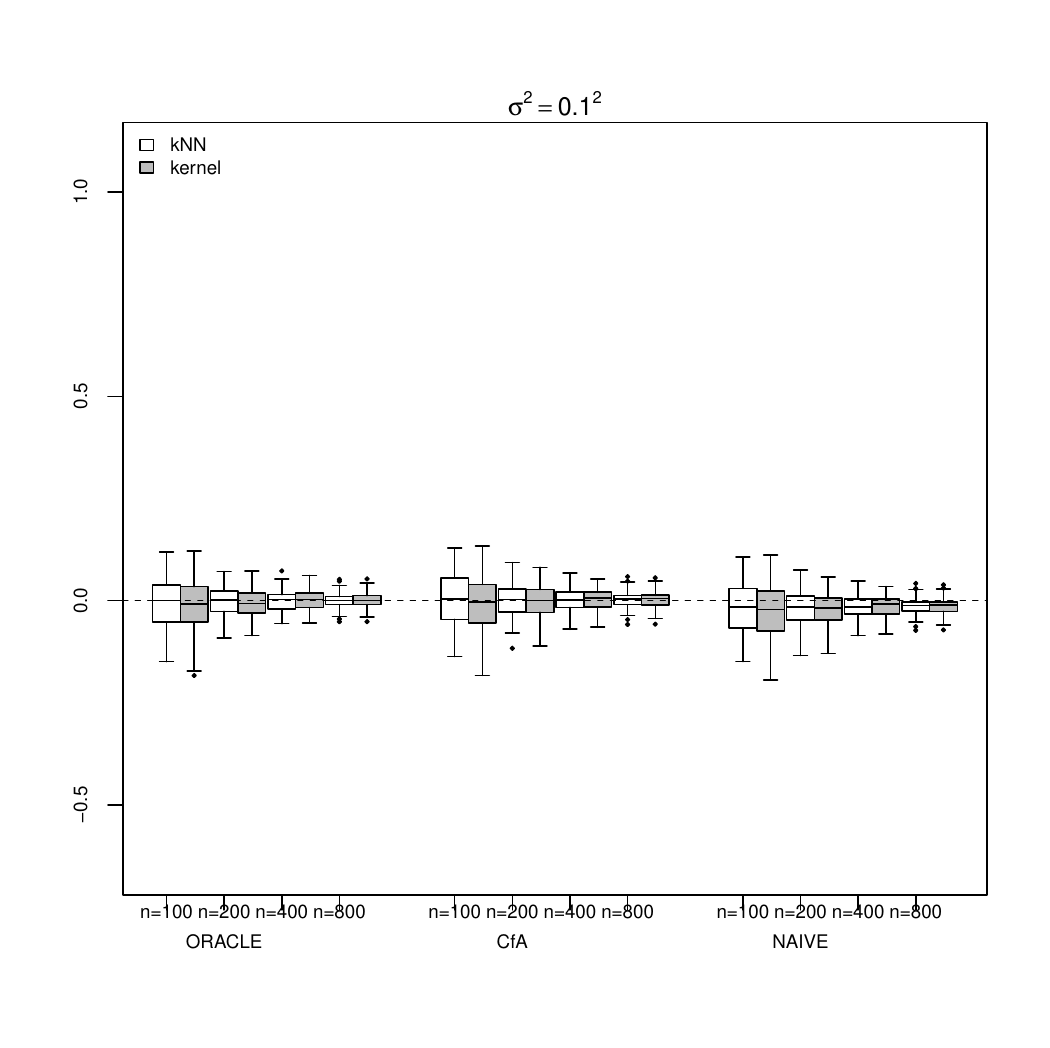} %
	\includegraphics[width=0.45\textwidth]{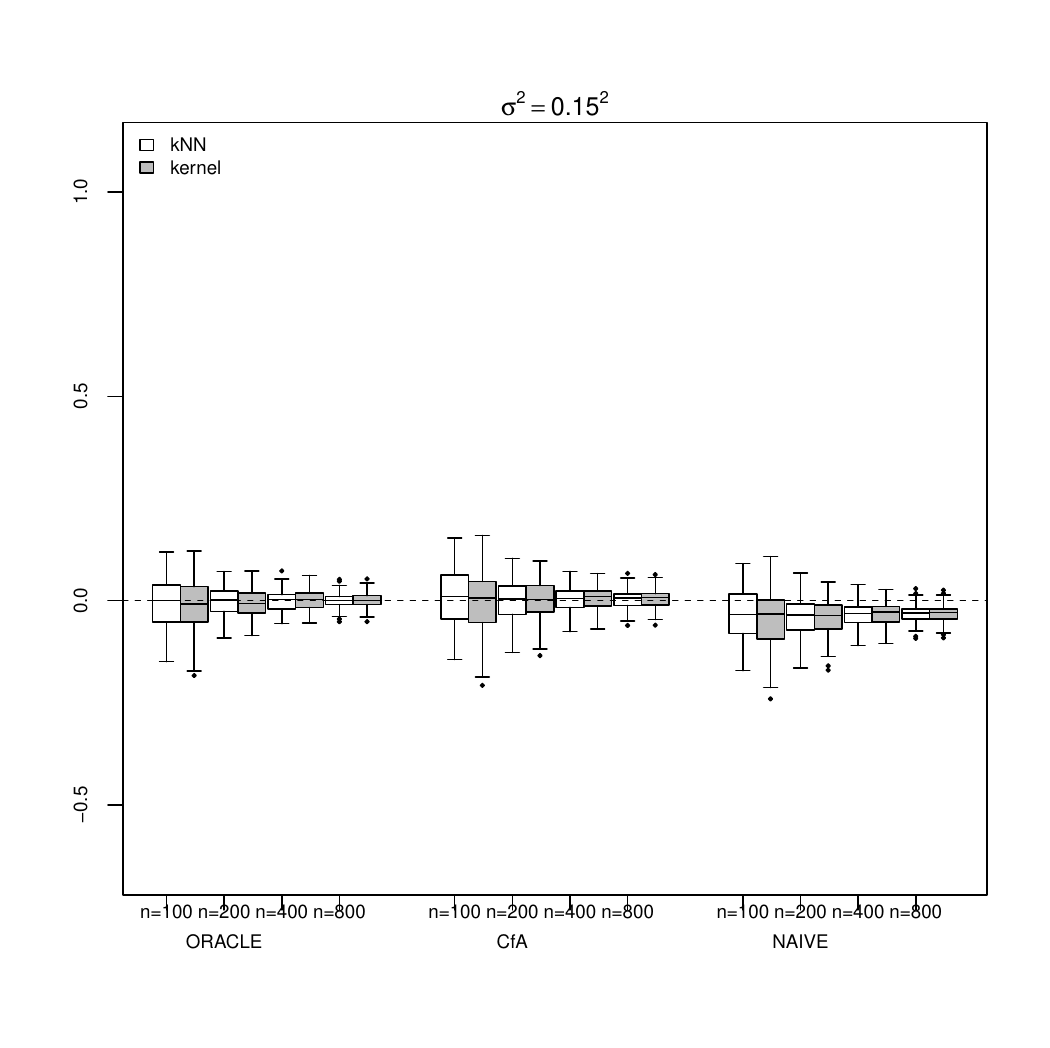}\newline
	\includegraphics[width=0.45\textwidth]{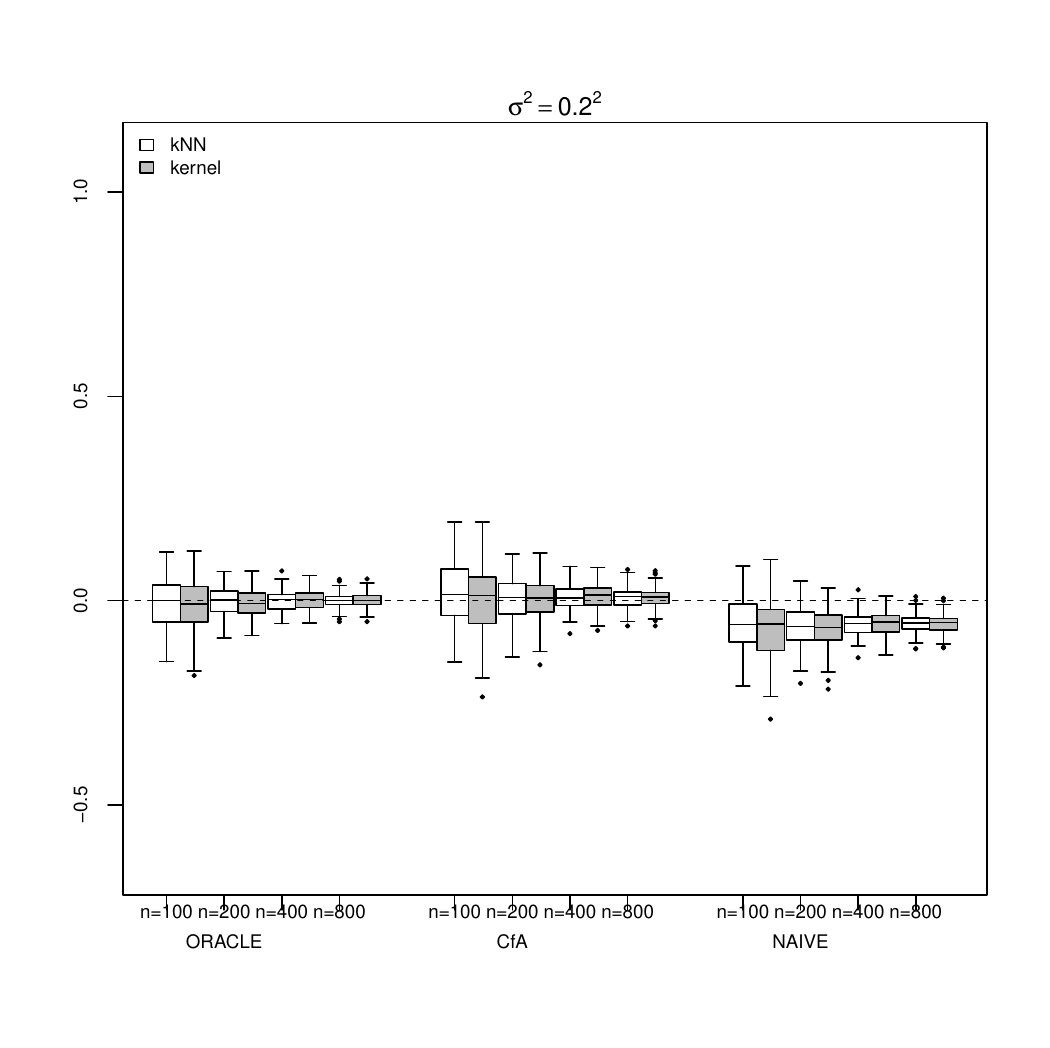} %
	\includegraphics[width=0.45\textwidth]{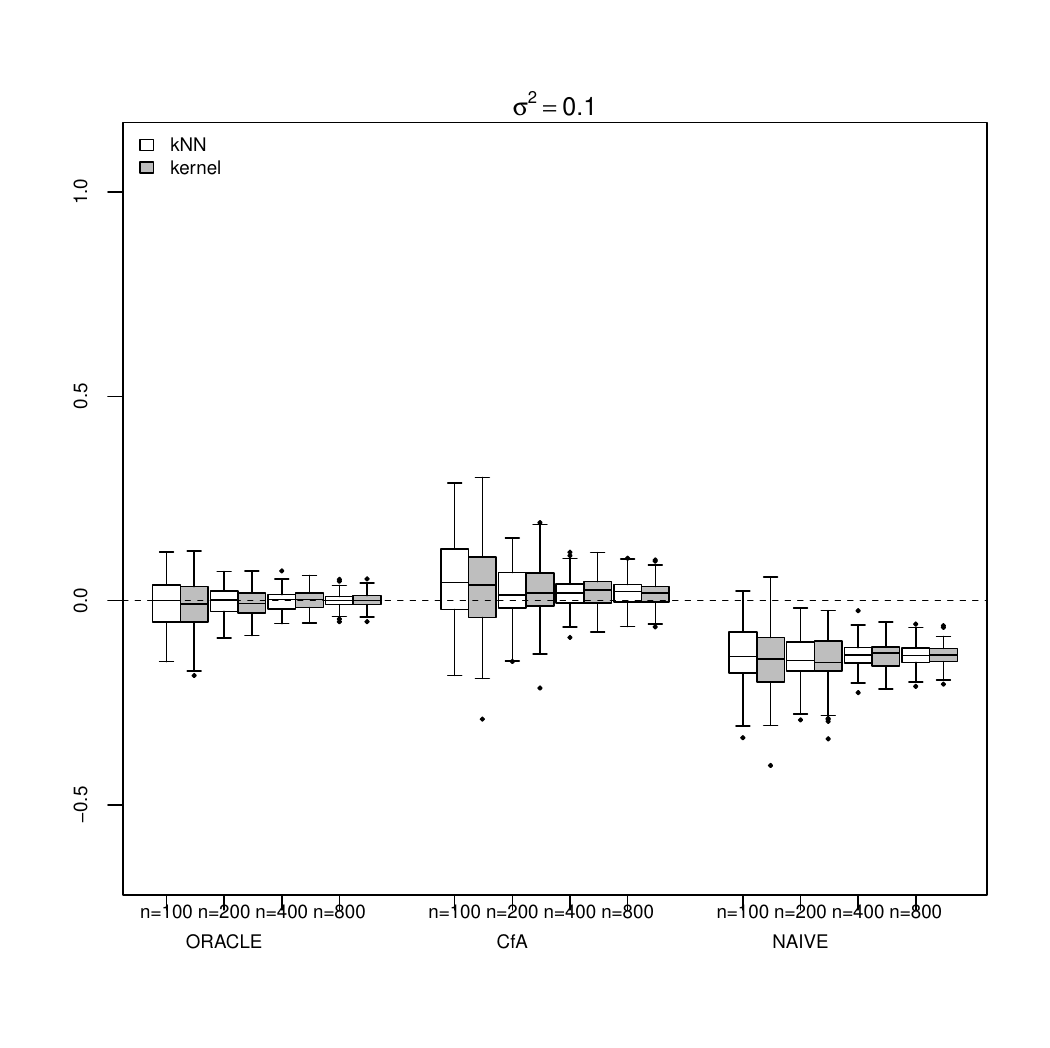}\newline
	\includegraphics[width=0.45\textwidth]{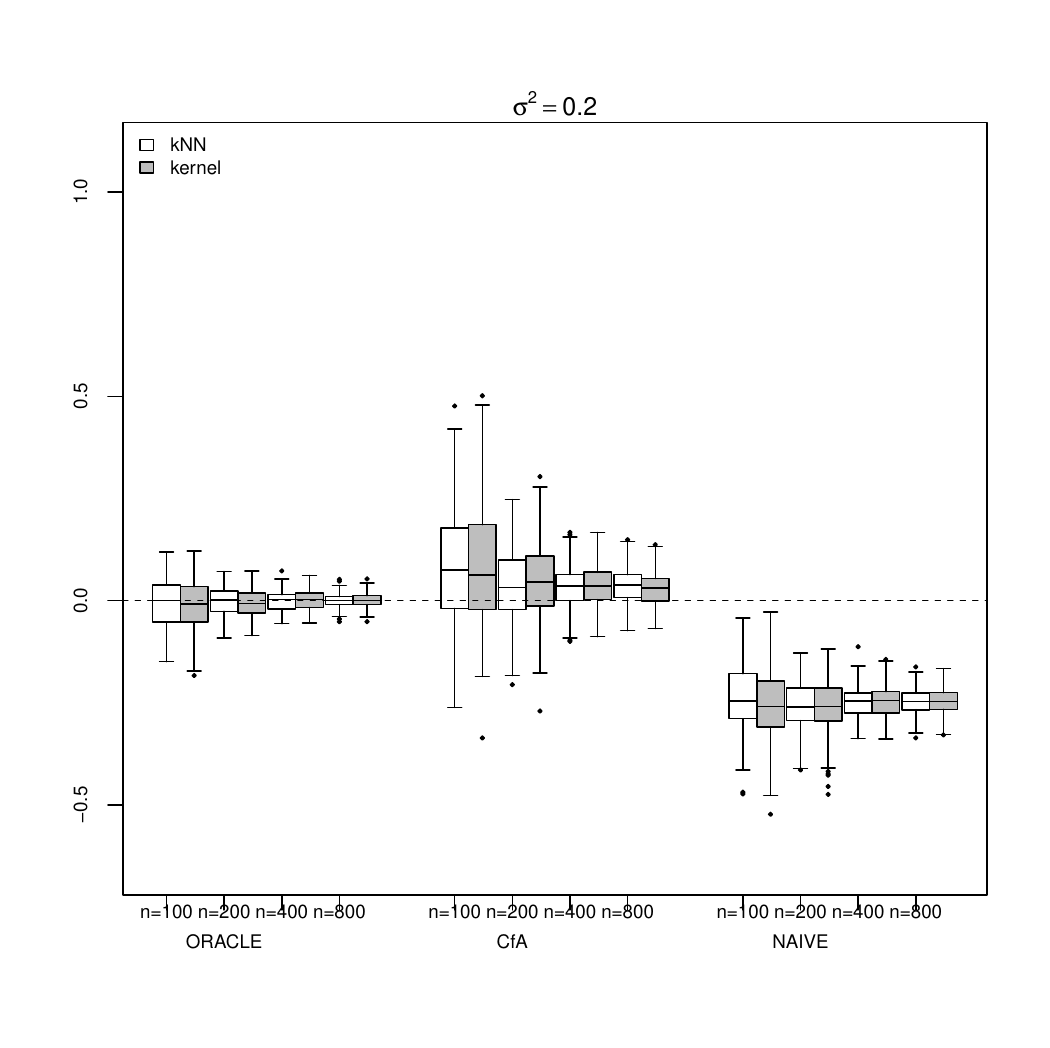} %
	\includegraphics[width=0.45\textwidth]{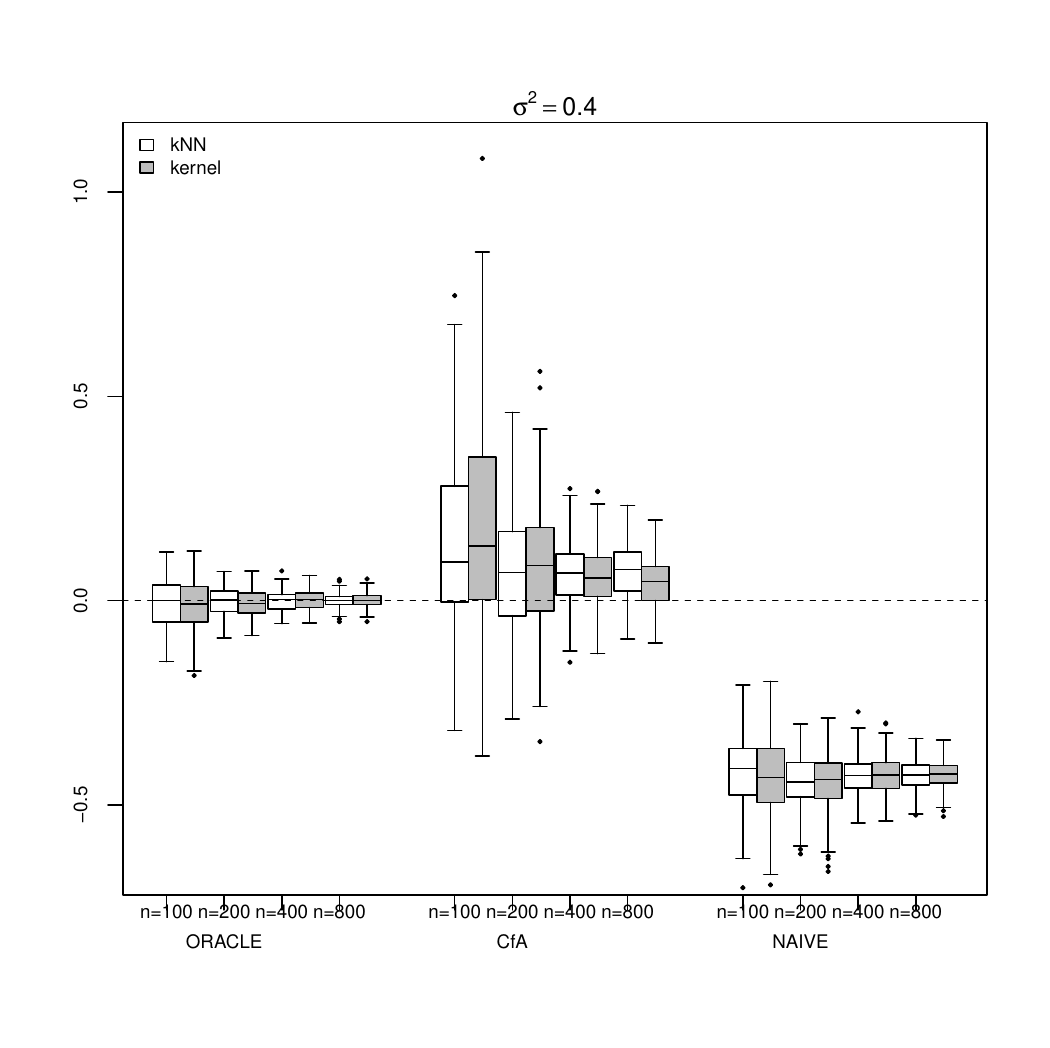}\newline
	
	\vspace{-1cm}
	\caption{Boxplots of $\overline{{\beta}}_{02}-\beta_{02}$ when ORACLE ($\overline{\pmb{\beta}}_{0}=\widetilde{\pmb{\beta}}_{0}$), CfA ($\overline{\pmb{\beta}}_{0}=\widehat{\pmb{\beta}}_{0}$) and NAIVE ($\overline{\pmb{\beta}}_{0}=\breve{\pmb{\beta}}_{0}$) procedures were used. For each of these three classes, both $k$NN- and kernel-based estimators were considered. Different sample sizes, $n$, and covariance matrices of the measurement error, $\pmb{\Sigma}_{uu}=\text{diag}(\sigma^2,\sigma^2)$, were used.}
	\label{fig2}
\end{figure}

Figures \ref{fig1} and \ref{fig2} show boxplots of $\overline{{\beta}}_{01}-\beta_{01}$ and $\overline{{\beta}}_{02}-\beta_{02}$, respectively, from the six different estimators, $\overline{\pmb{\beta}}_{0}=(\overline{{\beta}}_{01},\overline{{\beta}}_{02})^{\top}$, of $\pmb{\beta}_0=(\beta_{01},\beta_{02})^{\top}$ considered ($\overline{\pmb{\beta}}_{0}=\widetilde{\pmb{\beta}}_{0k},\widetilde{\pmb{\beta}}_{0h},\widehat{\pmb{\beta}}_{0k},\widehat{\pmb{\beta}}_{0h},\breve{\pmb{\beta}}_{0k},\breve{\pmb{\beta}}_{0h}$).

Figures \ref{fig1} and \ref{fig2} provide a comprehensive overview of the behaviour of the considered estimators. In particular, it is noticeable that our method (CfA) effectively reduces estimation bias. 
In contrast, the NAIVE procedure shows obvious biases. In all the considered estimators, variance decreases as sample size increases. As expected, the ORACLE estimator shows the best performance, a result of its use of uncontaminated information unlike the other estimators. In particular,  no bias is apparent, especially for large sample sizes. Comparing now the behaviour of $k$NN- and kernel-based estimators, in general their performances are more and more similar as the sample size increases. Finally, focusing on the CfA and NAIVE methods, it is evident that the performance worsens as the measurement error increases. All these comments are consistent with the expected.

Table 1 summarizes the MSE values of $\overline{{\beta}}_{0j}$ ($j=1,2$) obtained from the different estimators considered.
\begin{table}
	\scalebox{0.9}{
\begin{tabular}{p{0.6in}p{0.35in}p{0.35in}p{0.35in}p{0.35in}p{0in}p{0.35in}p{0.35in}p{0.35in}p{0.35in}p{0in}p{0.35in}p{0.35in}p{0.35in}p{0.35in}}
	\multicolumn{15}{l}{\textbf{Table 1}} \\ 
	\multicolumn{15}{l}{MSE of the estimators for $\pmb{\beta}_{0}$. For each estimator, the first and second lines correspond to the $k$NN and kernel} \\
	\multicolumn{15}{l}{versions, respectively.} \\ \hline
	$n$ & \multicolumn{1}{c}{$100$} & \multicolumn{1}{c}{$200$} & 
	\multicolumn{1}{c}{$400$} & \multicolumn{1}{c}{$800$} &  & 
	\multicolumn{1}{c}{$100$} & \multicolumn{1}{c}{$200$} & \multicolumn{1}{r}{$%
		400$} & \multicolumn{1}{c}{$800$} &  & \multicolumn{1}{c}{$100$} & 
	\multicolumn{1}{c}{$200$} & \multicolumn{1}{c}{$400$} & \multicolumn{1}{c}{$%
		800$} \\ \cline{1-15}
	& \multicolumn{1}{c}{} & \multicolumn{1}{c}{} & \multicolumn{1}{c}{} & 
	\multicolumn{1}{c}{} &  & \multicolumn{1}{c}{} & \multicolumn{1}{c}{} & 
	\multicolumn{1}{c}{} & \multicolumn{1}{c}{} &  & \multicolumn{1}{c}{} & 
	\multicolumn{1}{c}{} & \multicolumn{1}{c}{} & \multicolumn{1}{c}{} \\ 
	$\pmb{\Sigma}_{uu}$ & \multicolumn{4}{c}{$\text{diag}(0.1^{2},0.1^{2})$} & 
	& \multicolumn{4}{c}{$\text{diag}(0.15^{2},0.15^{2})$} &  & 
	\multicolumn{4}{c}{$\text{diag}(0.2^{2},0.2^{2})$} \\ 
	\cline{1-15}
	MSE$(\breve{\beta}_{1})$ & $%
	\begin{array}{l}
		0.0045 \\ 
		0.0056%
	\end{array}%
	$ & $%
	\begin{array}{l}
		0.0017 \\ 
		0.0017%
	\end{array}%
	$ & $%
	\begin{array}{l}
		0.0009 \\ 
		0.0008%
	\end{array}%
	$ & $%
	\begin{array}{l}
		0.0005 \\ 
		0.0005%
	\end{array}%
	$ &  & $%
	\begin{array}{l}
		0.0056 \\ 
		0.0068%
	\end{array}%
	$ & $%
	\begin{array}{l}
		0.0021 \\ 
		0.0023%
	\end{array}%
	$ & $%
	\begin{array}{l}
		0.0014 \\ 
		0.0013%
	\end{array}%
	$ & $%
	\begin{array}{l}
		0.0009 \\ 
		0.0008%
	\end{array}%
	$ &  & $%
	\begin{array}{l}
		0.0072 \\ 
		0.0086%
	\end{array}%
	$ & $%
	\begin{array}{l}
		0.0032 \\ 
		0.0035%
	\end{array}%
	$ & $%
	\begin{array}{l}
		0.0025 \\ 
		0.0024%
	\end{array}%
	$ & $%
	\begin{array}{l}
		0.0019 \\ 
		0.0019%
	\end{array}%
	$ \\ \cline{2-15}
	MSE$(\breve{\beta}_{2})$ & $%
	\begin{array}{l}
		0.0042 \\ 
		0.0055%
	\end{array}%
	$ & $%
	\begin{array}{l}
		0.0019 \\ 
		0.0020%
	\end{array}%
	$ & $%
	\begin{array}{l}
		0.0009 \\ 
		0.0009%
	\end{array}%
	$ & $%
	\begin{array}{l}
		0.0006 \\ 
		0.0006%
	\end{array}%
	$ &  & $%
	\begin{array}{l}
		0.0053 \\ 
		0.0072%
	\end{array}%
	$ & $%
	\begin{array}{l}
		0.0032 \\ 
		0.0035%
	\end{array}%
	$ & $%
	\begin{array}{l}
		0.0018 \\ 
		0.0017%
	\end{array}%
	$ & $%
	\begin{array}{l}
		0.0015 \\ 
		0.0015%
	\end{array}%
	$ &  & $%
	\begin{array}{l}
		0.0079 \\ 
		0.0104%
	\end{array}%
	$ & $%
	\begin{array}{l}
		0.0062 \\ 
		0.0068%
	\end{array}%
	$ & $%
	\begin{array}{l}
		0.0041 \\ 
		0.0040%
	\end{array}%
	$ & $%
	\begin{array}{l}
		0.0038 \\ 
		0.0037%
	\end{array}%
	$ \\ \hline
	MSE$(\widehat{\beta }_{1})$ & $%
	\begin{array}{l}
		0.0046 \\ 
		0.0056%
	\end{array}%
	$ & $%
	\begin{array}{l}
		0.0016 \\ 
		0.0016%
	\end{array}%
	$ & $%
	\begin{array}{l}
		0.0007 \\ 
		0.0007%
	\end{array}%
	$ & $%
	\begin{array}{l}
		0.0004 \\ 
		0.0004%
	\end{array}%
	$ &  & $%
	\begin{array}{l}
		0.0055 \\ 
		0.0067%
	\end{array}%
	$ & $%
	\begin{array}{l}
		0.0019 \\ 
		0.0019%
	\end{array}%
	$ & $%
	\begin{array}{l}
		0.0009 \\ 
		0.0009%
	\end{array}%
	$ & $%
	\begin{array}{l}
		0.0004 \\ 
		0.0004%
	\end{array}%
	$ &  & $%
	\begin{array}{l}
		0.0071 \\ 
		0.0083%
	\end{array}%
	$ & $%
	\begin{array}{l}
		0.0024 \\ 
		0.0023%
	\end{array}%
	$ & $%
	\begin{array}{l}
		0.0011 \\ 
		0.0011%
	\end{array}%
	$ & $%
	\begin{array}{l}
		0.0005 \\ 
		0.0005%
	\end{array}%
	$ \\ \cline{2-15}
	MSE$(\widehat{\beta }_{2})$ & $%
	\begin{array}{l}
		0.0042 \\ 
		0.0052%
	\end{array}%
	$ & $%
	\begin{array}{l}
		0.0016 \\ 
		0.0016%
	\end{array}%
	$ & $%
	\begin{array}{l}
		0.0007 \\ 
		0.0007%
	\end{array}%
	$ & $%
	\begin{array}{l}
		0.0004 \\ 
		0.0004%
	\end{array}%
	$ &  & $%
	\begin{array}{l}
		0.0048 \\ 
		0.0061%
	\end{array}%
	$ & $%
	\begin{array}{l}
		0.0019 \\ 
		0.0020%
	\end{array}%
	$ & $%
	\begin{array}{l}
		0.0008 \\ 
		0.0008%
	\end{array}%
	$ & $%
	\begin{array}{l}
		0.0005 \\ 
		0.0005%
	\end{array}%
	$ &  & $%
	\begin{array}{l}
		0.0064 \\ 
		0.0075%
	\end{array}%
	$ & $%
	\begin{array}{l}
		0.0025 \\ 
		0.0027%
	\end{array}%
	$ & $%
	\begin{array}{l}
		0.0010 \\ 
		0.0011%
	\end{array}%
	$ & $%
	\begin{array}{l}
		0.0007 \\ 
		0.0006%
	\end{array}%
	$ \\ \hline
	MSE$(\widetilde{\beta }_{1})$ & $%
	\begin{array}{l}
		0.0039 \\ 
		0.0045%
	\end{array}%
	$ & $%
	\begin{array}{l}
		0.0014 \\ 
		0.0014%
	\end{array}%
	$ & $%
	\begin{array}{l}
		0.0007 \\ 
		0.0007%
	\end{array}%
	$ & $%
	\begin{array}{l}
		0.0004 \\ 
		0.0003%
	\end{array}%
	$ &  & $%
	\begin{array}{l}
		0.0039 \\ 
		0.0045%
	\end{array}%
	$ & $%
	\begin{array}{l}
		0.0014 \\ 
		0.0014%
	\end{array}%
	$ & $%
	\begin{array}{l}
		0.0007 \\ 
		0.0007%
	\end{array}%
	$ & $%
	\begin{array}{l}
		0.0004 \\ 
		0.0003%
	\end{array}%
	$ &  & $%
	\begin{array}{l}
		0.0039 \\ 
		0.0045%
	\end{array}%
	$ & $%
	\begin{array}{l}
		0.0014 \\ 
		0.0014%
	\end{array}%
	$ & $%
	\begin{array}{l}
		0.0007 \\ 
		0.0007%
	\end{array}%
	$ & $%
	\begin{array}{l}
		0.0004 \\ 
		0.0003%
	\end{array}%
	$ \\ \cline{2-15}
	MSE$(\widetilde{\beta }_{2})$ & $%
	\begin{array}{l}
		0.0036 \\ 
		0.0043%
	\end{array}%
	$ & $%
	\begin{array}{l}
		0.0013 \\ 
		0.0013%
	\end{array}%
	$ & $%
	\begin{array}{l}
		0.0006 \\ 
		0.0006%
	\end{array}%
	$ & $%
	\begin{array}{l}
		0.0003 \\ 
		0.0004%
	\end{array}%
	$ &  & $%
	\begin{array}{l}
		0.0036 \\ 
		0.0043%
	\end{array}%
	$ & $%
	\begin{array}{l}
		0.0013 \\ 
		0.0013%
	\end{array}%
	$ & $%
	\begin{array}{l}
		0.0006 \\ 
		0.0006%
	\end{array}%
	$ & $%
	\begin{array}{l}
		0.0003 \\ 
		0.0004%
	\end{array}%
	$ &  & $%
	\begin{array}{l}
		0.0036 \\ 
		0.0043%
	\end{array}%
	$ & $%
	\begin{array}{l}
		0.0013 \\ 
		0.0013%
	\end{array}%
	$ & $%
	\begin{array}{l}
		0.0006 \\ 
		0.0006%
	\end{array}%
	$ & $%
	\begin{array}{l}
		0.0003 \\ 
		0.0004%
	\end{array}%
	$ \\ \hline 
	& \multicolumn{1}{r}{} & \multicolumn{1}{r}{} & \multicolumn{1}{r}{} & 
	\multicolumn{1}{r}{} &  & \multicolumn{1}{r}{} & \multicolumn{1}{r}{} & 
	\multicolumn{1}{r}{} & \multicolumn{1}{r}{} &  & \multicolumn{1}{r}{} & 
	\multicolumn{1}{r}{} & \multicolumn{1}{r}{} & \multicolumn{1}{r}{} \\ 
	$\pmb{\Sigma}_{uu}$ & \multicolumn{4}{c}{$\text{diag}(0.1,0.1)$} &  & 
	\multicolumn{4}{c}{$\text{diag}(0.2,0.2)$} &  & \multicolumn{4}{c}{$\text{%
			diag}(0.4,0.4)$} \\ \cline{1-15}
	MSE$(\breve{\beta}_{1})$ & $%
	\begin{array}{l}
		0.0156 \\ 
		0.0171%
	\end{array}%
	$ & $%
	\begin{array}{l}
		0.0104 \\ 
		0.0110%
	\end{array}%
	$ & $%
	\begin{array}{l}
		0.0099 \\ 
		0.0094%
	\end{array}%
	$ & $%
	\begin{array}{l}
		0.0087 \\ 
		0.0086%
	\end{array}%
	$ &  & $%
	\begin{array}{l}
		0.0342 \\ 
		0.0374%
	\end{array}%
	$ & $%
	\begin{array}{l}
		0.0306 \\ 
		0.0318%
	\end{array}%
	$ & $%
	\begin{array}{l}
		0.0301 \\ 
		0.0293%
	\end{array}%
	$ & $%
	\begin{array}{l}
		0.0281 \\ 
		0.0278%
	\end{array}%
	$ &  & $%
	\begin{array}{l}
		0.0842 \\ 
		0.0892%
	\end{array}%
	$ & $%
	\begin{array}{l}
		0.0873 \\ 
		0.0890%
	\end{array}%
	$ & $%
	\begin{array}{l}
		0.0852 \\ 
		0.0839%
	\end{array}%
	$ & $%
	\begin{array}{l}
		0.0816 \\ 
		0.0810%
	\end{array}%
	$ \\ \cline{2-15}
	MSE$(\breve{\beta}_{2})$ & $%
	\begin{array}{l}
		0.0229 \\ 
		0.0276%
	\end{array}%
	$ & $%
	\begin{array}{l}
		0.0241 \\ 
		0.0254%
	\end{array}%
	$ & $%
	\begin{array}{l}
		0.0196 \\ 
		0.0193%
	\end{array}%
	$ & $%
	\begin{array}{l}
		0.0190 \\ 
		0.0188%
	\end{array}%
	$ &  & $%
	\begin{array}{l}
		0.0652 \\ 
		0.0722%
	\end{array}%
	$ & $%
	\begin{array}{l}
		0.0718 \\ 
		0.0746%
	\end{array}%
	$ & $%
	\begin{array}{l}
		0.0637 \\ 
		0.0632%
	\end{array}%
	$ & $%
	\begin{array}{l}
		0.0629 \\ 
		0.0625%
	\end{array}%
	$ &  & $%
	\begin{array}{l}
		0.1806 \\ 
		0.1918%
	\end{array}%
	$ & $%
	\begin{array}{l}
		0.1993 \\ 
		0.2044%
	\end{array}%
	$ & $%
	\begin{array}{l}
		0.1858 \\ 
		0.1849%
	\end{array}%
	$ & $%
	\begin{array}{l}
		0.1845 \\ 
		0.1837%
	\end{array}%
	$ \\ \hline 
	MSE$(\widehat{\beta }_{1})$ & $%
	\begin{array}{l}
		0.0130 \\ 
		0.0151%
	\end{array}%
	$ & $%
	\begin{array}{l}
		0.0041 \\ 
		0.0044%
	\end{array}%
	$ & $%
	\begin{array}{l}
		0.0018 \\ 
		0.0019%
	\end{array}%
	$ & $%
	\begin{array}{l}
		0.0009 \\ 
		0.0009%
	\end{array}%
	$ &  & $%
	\begin{array}{l}
		0.0264 \\ 
		0.0309%
	\end{array}%
	$ & $%
	\begin{array}{l}
		0.0077 \\ 
		0.0089%
	\end{array}%
	$ & $%
	\begin{array}{l}
		0.0035 \\ 
		0.0037%
	\end{array}%
	$ & $%
	\begin{array}{l}
		0.0019 \\ 
		0.0017%
	\end{array}%
	$ &  & $%
	\begin{array}{l}
		0.0630 \\ 
		0.0835%
	\end{array}%
	$ & $%
	\begin{array}{l}
		0.0192 \\ 
		0.0225%
	\end{array}%
	$ & $%
	\begin{array}{l}
		0.0089 \\ 
		0.0086%
	\end{array}%
	$ & $%
	\begin{array}{l}
		0.0054 \\ 
		0.0040%
	\end{array}%
	$ \\ \cline{2-15}
	MSE$(\widehat{\beta }_{2})$ & $%
	\begin{array}{l}
		0.0131 \\ 
		0.0142%
	\end{array}%
	$ & $%
	\begin{array}{l}
		0.0047 \\ 
		0.0060%
	\end{array}%
	$ & $%
	\begin{array}{l}
		0.0019 \\ 
		0.0022%
	\end{array}%
	$ & $%
	\begin{array}{l}
		0.0014 \\ 
		0.0012%
	\end{array}%
	$ &  & $%
	\begin{array}{l}
		0.0278 \\ 
		0.0326%
	\end{array}%
	$ & $%
	\begin{array}{l}
		0.0099 \\ 
		0.0129%
	\end{array}%
	$ & $%
	\begin{array}{l}
		0.0040 \\ 
		0.0043%
	\end{array}%
	$ & $%
	\begin{array}{l}
		0.0031 \\ 
		0.0024%
	\end{array}%
	$ &  & $%
	\begin{array}{l}
		0.0694 \\ 
		0.0995%
	\end{array}%
	$ & $%
	\begin{array}{l}
		0.0262 \\ 
		0.0336%
	\end{array}%
	$ & $%
	\begin{array}{l}
		0.0111 \\ 
		0.0095%
	\end{array}%
	$ & $%
	\begin{array}{l}
		0.0091 \\ 
		0.0054%
	\end{array}%
	$ \\ \hline 
	MSE$(\widetilde{\beta }_{1})$ & $%
	\begin{array}{l}
		0.0039 \\ 
		0.0045%
	\end{array}%
	$ & $%
	\begin{array}{l}
		0.0014 \\ 
		0.0014%
	\end{array}%
	$ & $%
	\begin{array}{l}
		0.0007 \\ 
		0.0007%
	\end{array}%
	$ & $%
	\begin{array}{l}
		0.0004 \\ 
		0.0003%
	\end{array}%
	$ &  & $%
	\begin{array}{l}
		0.0039 \\ 
		0.0045%
	\end{array}%
	$ & $%
	\begin{array}{l}
		0.0014 \\ 
		0.0014%
	\end{array}%
	$ & $%
	\begin{array}{l}
		0.0007 \\ 
		0.0007%
	\end{array}%
	$ & $%
	\begin{array}{l}
		0.0004 \\ 
		0.0003%
	\end{array}%
	$ &  & $%
	\begin{array}{l}
		0.0039 \\ 
		0.0045%
	\end{array}%
	$ & $%
	\begin{array}{l}
		0.0014 \\ 
		0.0014%
	\end{array}%
	$ & $%
	\begin{array}{l}
		0.0007 \\ 
		0.0007%
	\end{array}%
	$ & $%
	\begin{array}{l}
		0.0004 \\ 
		0.0003%
	\end{array}%
	$ \\ \cline{2-15}
	MSE$(\widetilde{\beta }_{2})$ & $%
	\begin{array}{l}
		0.0036 \\ 
		0.0043%
	\end{array}%
	$ & $%
	\begin{array}{l}
		0.0013 \\ 
		0.0013%
	\end{array}%
	$ & $%
	\begin{array}{l}
		0.0006 \\ 
		0.0006%
	\end{array}%
	$ & $%
	\begin{array}{l}
		0.0003 \\ 
		0.0004%
	\end{array}%
	$ &  & $%
	\begin{array}{l}
		0.0036 \\ 
		0.0043%
	\end{array}%
	$ & $%
	\begin{array}{l}
		0.0013 \\ 
		0.0013%
	\end{array}%
	$ & $%
	\begin{array}{l}
		0.0006 \\ 
		0.0006%
	\end{array}%
	$ & $%
	\begin{array}{l}
		0.0003 \\ 
		0.0004%
	\end{array}%
	$ &  & $%
	\begin{array}{l}
		0.0036 \\ 
		0.0043%
	\end{array}%
	$ & $%
	\begin{array}{l}
		0.0013 \\ 
		0.0013%
	\end{array}%
	$ & $%
	\begin{array}{l}
		0.0006 \\ 
		0.0006%
	\end{array}%
	$ & $%
	\begin{array}{l}
		0.0003 \\ 
		0.0004%
	\end{array}%
	$ \\ \hline
\end{tabular}}
\end{table}
\bigskip

\noindent It can be seen from Table 1 that our proposed procedure (CfA) for the parametric component outperforms the NAIVE approach, as the MSE of the CfA is smaller. Furthermore, as the measurement error increases, the difference between them increases and their performance also deteriorates. The MSEs for both CfA and ORACLE decrease with increasing sample size. In addition, in scenarios of low measurement error, the performance of the CfA approaches that of the ORACLE as the sample size increases;  as expected, the ORACLE estimator shows the best performance. Focusing now on a comparison between the $k$NN- and kernel-based estimators, it seems that for small sample sizes the $k$NN-based estimators outperform the kernel-based ones. In general, their behaviours are similar for large sample sizes. 

Table 2 reports the MSEP values obtained by the ORACLE (MSEP$(\tilde{Y})$), CfA (MSEP$(\widehat{Y})$) and NAIVE (MSEP$(\breve{Y})$) procedures.

\begin{table}
	\scalebox{0.9}{
\begin{tabular}{p{0.6in}p{0.35in}p{0.35in}p{0.35in}p{0.35in}cp{0.35in}p{0.35in}p{0.35in}p{0.35in}cp{0.35in}p{0.35in}p{0.35in}p{0.35in}}
	\multicolumn{15}{l}{\textbf{Table 2}} \\ 
	\multicolumn{15}{l}{MSEP from the different procedures considered. For each estimator, the first and second lines correspond to} \\
	\multicolumn{15}{l}{the $k$NN and kernel versions, respectively.} \\ \hline
	$n$ & \multicolumn{1}{c}{$100$} & \multicolumn{1}{c}{$200$} & 
	\multicolumn{1}{c}{$400$} & \multicolumn{1}{c}{$800$} &  & 
	\multicolumn{1}{c}{$100$} & \multicolumn{1}{c}{$200$} & \multicolumn{1}{c}{$%
		400$} & \multicolumn{1}{c}{$800$} &  & \multicolumn{1}{c}{$100$} & 
	\multicolumn{1}{c}{$200$} & \multicolumn{1}{c}{$400$} & \multicolumn{1}{c}{$%
		800$} \\ \hline 
	& \multicolumn{1}{c}{} & \multicolumn{1}{c}{} & \multicolumn{1}{c}{} & 
	\multicolumn{1}{c}{} &  & \multicolumn{1}{c}{} & \multicolumn{1}{c}{} & 
	\multicolumn{1}{c}{} & \multicolumn{1}{c}{} &  & \multicolumn{1}{c}{} & 
	\multicolumn{1}{c}{} & \multicolumn{1}{c}{} & \multicolumn{1}{c}{} \\ 
	$\pmb{\Sigma}_{uu}$ & \multicolumn{4}{c}{$\text{diag}(0.1^{2},0.1^{2})$} & 
	& \multicolumn{4}{c}{$\text{diag}(0.15^{2},0.15^{2})$} &  & 
	\multicolumn{4}{c}{$\text{diag}(0.2^{2},0.2^{2})$} \\ 
	\cline{1-15}
	MSEP$(\breve{Y})$ & $%
	\begin{array}{l}
		0.3836 \\ 
		0.6485%
	\end{array}%
	$ & $%
	\begin{array}{l}
		0.3129 \\ 
		0.3558%
	\end{array}%
	$ & $%
	\begin{array}{l}
		0.2778 \\ 
		0.2974%
	\end{array}%
	$ & $%
	\begin{array}{l}
		0.2696 \\ 
		0.2764%
	\end{array}%
	$ &  & $%
	\begin{array}{l}
		0.3961 \\ 
		0.6580%
	\end{array}%
	$ & $%
	\begin{array}{l}
		0.3192 \\ 
		0.3636%
	\end{array}%
	$ & $%
	\begin{array}{l}
		0.2822 \\ 
		0.3043%
	\end{array}%
	$ & $%
	\begin{array}{l}
		0.2735 \\ 
		0.2805%
	\end{array}%
	$ &  & $%
	\begin{array}{l}
		0.4132 \\ 
		0.6715%
	\end{array}%
	$ & $%
	\begin{array}{l}
		0.3286 \\ 
		0.3755%
	\end{array}%
	$ & $%
	\begin{array}{l}
		0.2896 \\ 
		0.3130%
	\end{array}%
	$ & $%
	\begin{array}{l}
		0.2801 \\ 
		0.2867%
	\end{array}%
	$ \\ \cline{2-15}
	MSEP$(\widehat{Y})$ & $%
	\begin{array}{l}
		0.3840 \\ 
		0.6486%
	\end{array}%
	$ & $%
	\begin{array}{l}
		0.3129 \\ 
		0.3559%
	\end{array}%
	$ & $%
	\begin{array}{l}
		0.2774 \\ 
		0.2975%
	\end{array}%
	$ & $%
	\begin{array}{l}
		0.2694 \\ 
		0.2761%
	\end{array}%
	$ &  & $%
	\begin{array}{l}
		0.3950 \\ 
		0.6585%
	\end{array}%
	$ & $%
	\begin{array}{l}
		0.3182 \\ 
		0.3625%
	\end{array}%
	$ & $%
	\begin{array}{l}
		0.2805 \\ 
		0.3028%
	\end{array}%
	$ & $%
	\begin{array}{l}
		0.2722 \\ 
		0.2788%
	\end{array}%
	$ &  & $%
	\begin{array}{l}
		0.4107 \\ 
		0.6723%
	\end{array}%
	$ & $%
	\begin{array}{l}
		0.3256 \\ 
		0.3724%
	\end{array}%
	$ & $%
	\begin{array}{l}
		0.2846 \\ 
		0.3092%
	\end{array}%
	$ & $%
	\begin{array}{l}
		0.2759 \\ 
		0.2823%
	\end{array}%
	$ \\ \cline{2-15}
	MSEP$(\widetilde{Y})$ & $%
	\begin{array}{l}
		0.3729 \\ 
		0.6407%
	\end{array}%
	$ & $%
	\begin{array}{l}
		0.3082 \\ 
		0.3512%
	\end{array}%
	$ & $%
	\begin{array}{l}
		0.2745 \\ 
		0.2928%
	\end{array}%
	$ & $%
	\begin{array}{l}
		0.2671 \\ 
		0.2736%
	\end{array}%
	$ &  & $%
	\begin{array}{l}
		0.3729 \\ 
		0.6407%
	\end{array}%
	$ & $%
	\begin{array}{l}
		0.3082 \\ 
		0.3512%
	\end{array}%
	$ & $%
	\begin{array}{l}
		0.2745 \\ 
		0.2928%
	\end{array}%
	$ & $%
	\begin{array}{l}
		0.2671 \\ 
		0.2736%
	\end{array}%
	$ &  & $%
	\begin{array}{l}
		0.3729 \\ 
		0.6407%
	\end{array}%
	$ & $%
	\begin{array}{l}
		0.3082 \\ 
		0.3512%
	\end{array}%
	$ & $%
	\begin{array}{l}
		0.2745 \\ 
		0.2928%
	\end{array}%
	$ & $%
	\begin{array}{l}
		0.2671 \\ 
		0.2736%
	\end{array}%
	$ \\ \cline{2-15}
	&  &  &  &  &  &  &  &  &  &  &  &  &  &  \\ 
	$\pmb{\Sigma}_{uu}$ & \multicolumn{4}{c}{$\text{diag}(0.1,0.1)$} &  & 
	\multicolumn{4}{c}{$\text{diag}(0.2,0.2)$} &  & \multicolumn{4}{c}{$\text{%
			diag}(0.4,0.4)$} \\ \cline{1-15}
	MSEP$(\breve{Y})$ & $%
	\begin{array}{l}
		0.4678 \\ 
		0.7238%
	\end{array}%
	$ & $%
	\begin{array}{l}
		0.3719 \\ 
		0.4278%
	\end{array}%
	$ & $%
	\begin{array}{l}
		0.3249 \\ 
		0.3545%
	\end{array}%
	$ & $%
	\begin{array}{l}
		0.3130 \\ 
		0.3192%
	\end{array}%
	$ &  & $%
	\begin{array}{l}
		0.5704 \\ 
		0.8282%
	\end{array}%
	$ & $%
	\begin{array}{l}
		0.4620 \\ 
		0.5328%
	\end{array}%
	$ & $%
	\begin{array}{l}
		0.4067 \\ 
		0.4426%
	\end{array}%
	$ & $%
	\begin{array}{l}
		0.3923 \\ 
		0.3962%
	\end{array}%
	$ &  & $%
	\begin{array}{l}
		0.7990 \\ 
		1.0599%
	\end{array}%
	$ & $%
	\begin{array}{l}
		0.6773 \\ 
		0.7710%
	\end{array}%
	$ & $%
	\begin{array}{l}
		0.6126 \\ 
		0.6537%
	\end{array}%
	$ & $%
	\begin{array}{l}
		0.5930 \\ 
		0.5911%
	\end{array}%
	$ \\ \cline{2-15}
	MSEP$(\widehat{Y})$ & $%
	\begin{array}{l}
		0.4569 \\ 
		0.7259%
	\end{array}%
	$ & $%
	\begin{array}{l}
		0.3489 \\ 
		0.4134%
	\end{array}%
	$ & $%
	\begin{array}{l}
		0.2993 \\ 
		0.3323%
	\end{array}%
	$ & $%
	\begin{array}{l}
		0.2895 \\ 
		0.2945%
	\end{array}%
	$ &  & $%
	\begin{array}{l}
		0.5518 \\ 
		0.8399%
	\end{array}%
	$ & $%
	\begin{array}{l}
		0.3858 \\ 
		0.4763%
	\end{array}%
	$ & $%
	\begin{array}{l}
		0.3244 \\ 
		0.3651%
	\end{array}%
	$ & $%
	\begin{array}{l}
		0.3137 \\ 
		0.3129%
	\end{array}%
	$ &  & $%
	\begin{array}{l}
		0.7734 \\ 
		1.1632%
	\end{array}%
	$ & $%
	\begin{array}{l}
		0.4616 \\ 
		0.6202%
	\end{array}%
	$ & $%
	\begin{array}{l}
		0.3839 \\ 
		0.4369%
	\end{array}%
	$ & $%
	\begin{array}{l}
		0.3693 \\ 
		0.3502%
	\end{array}%
	$ \\ \cline{2-15}
	MSEP$(\widetilde{Y})$ & $%
	\begin{array}{l}
		0.3729 \\ 
		0.6407%
	\end{array}%
	$ & $%
	\begin{array}{l}
		0.3082 \\ 
		0.3512%
	\end{array}%
	$ & $%
	\begin{array}{l}
		0.2745 \\ 
		0.2928%
	\end{array}%
	$ & $%
	\begin{array}{l}
		0.2671 \\ 
		0.2736%
	\end{array}%
	$ &  & $%
	\begin{array}{l}
		0.3729 \\ 
		0.6407%
	\end{array}%
	$ & $%
	\begin{array}{l}
		0.3082 \\ 
		0.3512%
	\end{array}%
	$ & $%
	\begin{array}{l}
		0.2745 \\ 
		0.2928%
	\end{array}%
	$ & $%
	\begin{array}{l}
		0.2671 \\ 
		0.2736%
	\end{array}%
	$ &  & $%
	\begin{array}{l}
		0.3729 \\ 
		0.6407%
	\end{array}%
	$ & $%
	\begin{array}{l}
		0.3082 \\ 
		0.3512%
	\end{array}%
	$ & $%
	\begin{array}{l}
		0.2745 \\ 
		0.2928%
	\end{array}%
	$ & $%
	\begin{array}{l}
		0.2671 \\ 
		0.2736%
	\end{array}%
	$ \\ \hline
\end{tabular}}
\end{table}
\bigskip
\noindent It can be seen from Table 2 that, when the sample size is small, the predictive performance of each $k$NN-based procedure is better than that of the corresponding kernel-based method. In addition, as the sample size increases their predictive performances become more similar. Focusing now on the type of method considered (ORACLE, CfA or NAIVE), it can be noted that while for scenarios of low measurement error the predictive performance of the three types of methods is close, the differences between their preformances increase as the measurement error increases. As expected, ORACLE and NAIVE show the best and worst performances, respectively, while CfA exhibits competitive behaviour in any scenario of measurement error. Notably, the performance of the NAIVE method in scenarios with moderate to high measurement error is particularly poor.

\section{Real data analysis}\label{reald}

This section is devoted to comparing, on a real data set, the predictive power of the SFPLME model (\ref{eq_model1})-(\ref{EM}) when estimated by the proposed $k$NN-based estimators (\ref{beta-est}) and (\ref{kNN_est}) against the case where the kernel-based estimators proposed in \cite{zhu20} are considered. The real data set is the well-known Tecator's data set, which is a benchmark data set in the setting of FDA (for the particular case of error-in-variables models, see, for instance, \citealt{zhu19} and \citealt{zhu20}).

Tecators's data include the percentages of fat, protein and moisture contents, as well as the near-infrared absorbance spectra of $215$ finely chopped pieces of meat. For each piece of meat, the percentages of fat, protein and moisture ($Y_i$, $X_{i1}$ and $X_{i2}$, respectively) are scalar, while the corresponding near-infrared absorbance spectra observations were collected at $100$ equally spaced wavelengths ($t_j$, $j=1,\dots,100$) in the range $850$--$1050$ $nm$; so each subject can be considered as a continuous curve, $\mathcal{X}_i$.

Firstly, we split the original sample into two subsamples: a training sample, $$\mathcal{S}_{train.1}=\{(X_{i1},X_{i2},\mathcal{X}_i,Y_i)\}_{i=1}^{n},$$ and a testing one, $$\mathcal{S}_{test}=\{(X_{i1},X_{i2},\mathcal{X}_i,Y_i)\}_{i=n+1}^{215}.$$ 
We constructed a second training sample, $\mathcal{S}_{train.2}$, by adding measurement errors to the covariate $X_{i}=(X_{i1},X_{i2})^{\top}$ ($i=1,\ldots,n$); specifically, $$\mathcal{S}_{train.2}=\{(W_{i1},W_{i2},\mathcal{X}_i,Y_i)\}_{i=1}^{n},$$ where
$$W_{i}=X_{i}+U_{i}, \ (i=1,\ldots,n)$$ with $U_{i}=(U_{i1},U_{i2})^{\top} \sim N(\pmb{0},\pmb{\Sigma}_{uu})$ (we have denoted $W_{i}=(W_{i1},W_{i2})^{\top}, \ i=1,\ldots,n$).

In this real data application, we assumed that $\pmb{\Sigma}_{uu}$ is unknown and we estimated it by means of $\widehat{\pmb{\Sigma}}_{uu}$ (see (\ref{Sigma_est}) in Remark \ref{unknown}). Then, the corresponding CfA-$k$NN-based estimators of $\pmb{\beta}_{0}$ and $m(\chi)$ are $\widehat{\pmb{\beta}}_{0k}^{\ast}$ and $\widehat{m}_{k}(\chi)^{\ast}$, respectively (see (\ref{beta-est-replic}) and (\ref{kNN_est}), respectively). The CfA-kernel-based estimators of $\pmb{\beta}_{0}$ and $m(\cdot)$ differs from the CfA-$k$NN-based ones only in the use of weights (\ref{pesos-kernel}) instead of (\ref{pesos}). 

The training samples $\mathcal{S}_{train.1}$ and $\mathcal{S}_{train.2}$ (including $r$ replicates of $W_i$) were used to obtain the ORACLE and CfA estimators, respectively, including the tuning parameters $k$, $h$ and $q$; the testing sample was employed to measure the quality of the predictions using the MSEP. The same kernel and class of semi-metrics as in Section \ref{simul} were considered.

Values $n=165$ and $r=2$ were used. Regarding the covariance matrices of the measurement errors, $\pmb{\Sigma}_{uu}$, the matrices diag$(0.5^2,0.5^2)$, diag$(1^2,1^2)$, diag$(1.5^2,1.5^2)$ and diag$(2^2,2^2)$ were considered. The experiment was repeated $M=200$ times.
Average of the corresponding MSEP values are displayed in Table 3.

\bigskip
\begin{tabular}{cccccccc}
	\multicolumn{8}{l}{\textbf{Table 3}} \\ 
	\multicolumn{8}{l}{MSEP obtained from CfA-$k$NN- and CfA-kernel-based procedures.} \\ \hline
	$\pmb{\Sigma}_{uu}$ & $\text{diag}(0.5^{2},0.5^{2})$ &  & $\text{diag}%
	(1^{2},1^{2})$ &  & $\text{diag}(1.5^{2},1.5^{2})$ &  & $\text{diag}%
	(2^{2},2^{2})$ \\ \cline{1-8}
	\multicolumn{1}{l}{%
		\begin{tabular}{l}
			$k$NN \\ 
			kernel%
		\end{tabular}%
	} & $%
	\begin{array}{l}
		0.6135 \\ 
		0.8872%
	\end{array}%
	$ &  & $%
	\begin{array}{l}
		0.8343 \\ 
		0.9511%
	\end{array}%
	$ &  & $%
	\begin{array}{l}
		1.1144 \\ 
		1.1663%
	\end{array}%
	$ &  & $%
	\begin{array}{l}
		1.5339 \\ 
		1.5743%
	\end{array}%
	$ \\ \hline
\end{tabular}

\bigskip
\noindent Table 3 shows that the predictive power of the SFPLME model is enhanced when employing our proposal (CfA-$k$NN-based estimators) compared to using the approach in \cite{zhu20} (CfA-kernel-based estimators). As anticipated, the performance of both methods declines as the variance of the measurement errors increases.

Finally, it is important to note that, to the best of our knowledge, \cite{zhu19} is the only paper in the statistical literature on error-in-variables models that illustrates predictive performance using the Tecator's data set. Specifically, \cite{zhu19} considered the partially functional linear error-in-variables model $$Y=X^{\top}\pmb{\beta}_0 + \int \alpha(t)\mathcal{X}(t)dt + \varepsilon,$$ where one observes $W=X+U$ instead of $X.$ In a similar analysis as the one shown here, the value of the MSEP they obtained when $\pmb{\Sigma}_{uu}=\text{diag}(2^{2},2^{2})$ was $1.9684$. This value supports the suitability of the SFPLME to model the Tecator's data set when error in variables is introduced.

\section{Concluding comments}\label{conclu}

This paper has stated some first asymptotic theory of $k$NN estimation in the semi-functional partial linear regression model with measurement errors in the covariates with linear effect.  It was shown that the impact of considering measurement errors involves modifying the covariance matrix of the asymptotic distribution of the estimator for the parameter vector in the linear component of the model (consequently, the corresponding limit in the law of the iterated logarithm is also modified). 
Measurement errors have no impact on the rates of uniform convergence of the estimator of the nonparametric component. To obtain the proofs of the main theoretical results, it was necessary to prove functional versions of some lemmas presented in \cite{lia99}; for instance, our Lemma \ref{lemma6.5} (see Appendix \ref{app}) extends Lemma A.7 in \cite{lia99} from the scalar setting to the functional one (in addition, while our lemmas consider almost sure convergence, the convergence in the lemmas from \citealt{lia99} is in probability). It should also be noted that our Lemma \ref{lemma6.5} extends Lemma \ref{lemma3} (see Appendix \ref{app}) to the setting of measurement errors (Lemma \ref{lemma3} was proved and used in \citealt{linav20}, where the SFPL model without measurement errors (\ref{eq_model1}) was addressed). Furthermore, some results already existing in the literature were also used. The finite sample size study highlighted the effectiveness of the proposed estimator for the parametric component in reducing the estimation bias. Moreover, it demonstrated superior performance of this estimator compared to the one in \cite{zhu20} (based on kernel estimation) especially in the context of small sample sizes. This superiority of the $k$NN-based proposal against the kernel-based one was also observed in an application to real data.  Extensions to other kinds of semiparametric models are challenging open problems for which the technique used in this paper could be helpful.

\bigskip
\noindent{\large \bf{Acknowledgements}}
\noindent This research/work is part of the grants PID2020-113578RB-I00 and PID2023-147127OB-I00 ``ERDF/EU'', funded by MCIN/AEI/10.13039/501100011033/. It has also been supported by the Xunta de Galicia (Grupos de Referencia Competitiva ED431C-2024/14) and by CITIC as a center accredited for excellence within the Galician University System and a member of the CIGUS Network, receives subsidies from the Department of Education, Science, Universities, and Vocational Training of the Xunta de Galicia. Additionally, it is co-financed by the EU through the FEDER Galicia 2021-27 operational program (Ref. ED431G 2023/01). The authors thank two anonymous referees for their constructive comments and suggestions which helped to improve the quality of this paper.

\appendix

\section{Appendix: Proofs}\label{app}

First of all, let us remember that, for any $(n\times q)$-matrix $\mathbf {A}$ $(q\geq 1)$ and number of neighbours $k$, we denote
$\widetilde{\mathbf {A}}=\left(\mathbf {I}-\mathbf {V}_{k}\right)\mathbf {A}, \mbox{ where } \mathbf {V}_{k}=\left(\omega_{k}(\mathcal{X}_i,\mathcal{X}_j)\right)_{i,j}$. In addition, the $i$th row of $\widetilde{\mathbf {A}}$ will be denoted by $\widetilde{A}_i$ while the $j$th component of $\widetilde{A}_i$ will be denoted by $\widetilde{A}_{ij}$ ($i=1,\ldots,n, \ j=1,\ldots,q$); that is, $\widetilde{\mathbf {A}}=(\widetilde{A}_1,\ldots,\widetilde{A}_n)^{\top}=\left(\widetilde{A}_{ij}\right).$ Finally, we denote $\pmb{\eta}=(\eta_1,\ldots,\eta_n)^{\top}$, $\pmb{\varepsilon}=(\varepsilon_1, \ldots,\varepsilon_n)^{\top}$, $\pmb{m}=(m(\mathcal{X}_1),\ldots, m(\mathcal{X}_n))^{\top}$ and $\mathbf{G}=(G_1,\ldots,G_n)^{\top}$ with $G_i=(g_1(\mathcal{X}_i),\ldots,g_p(\mathcal{X}_i))^{\top}$ ($i=1,\ldots,n$).

In this Appendix, we first present some technical lemmas to be used in the proofs of our theorems. Some of these lemmas are known while the others are slight modifications of known lemmas. Then, we obtain the proofs of our main results (Theorem \ref{theorem1}).

\subsection{Technical lemmas}\label{app1}

\begin{lemma}
	\label{LemaStout} (\citealt{stout74}, Corollary 5.2.3) Let $V_{1},...,V_{n}$ be independent r.v. with 0 means and
	$\max_{1\leq i\leq n}E\left\vert V_{i}\right\vert ^{2+\delta}<\infty$, for
	some $\delta>0$. If, in addition, $\lim\inf_{n\rightarrow\infty}n^{-1}\sum_{i=1}^{n}Var\left(
	V_{i}\right)  >0,$ then%
	\[
	\lim\sup_{n\rightarrow\infty}\left\vert S_{n}\right\vert /\left(  2s_{n}%
	^{2}\log\log s_{n}^{2}\right)  ^{1/2}=1\text{ a.s.,}%
	\]
	where $S_{n}=\sum_{i=1}^{n}V_{i}$ and $s_{n}^{2}=\sum_{i=1}^{n}Var\left(
	V_{i}\right)  $.
\end{lemma}

\label{lemmas}
\begin{lemma}\label{lemma1} (\citealt{Kud13}, Theorem 2) If assumptions (A1)-(A6)(i) are satisfied then we have that\\
	\begin{equation}
		\sup_{\chi\in S_{\mathcal{F}}}\left|g_{j}(\chi)-\sum_{i=1}^{n}\omega_{k}(\chi,\mathcal{X}_i)X_{ij} \right|= O\left(\phi^{-1}\left(\frac{k}{n}\right) ^{\alpha}+\sqrt{\frac{\psi_{S_{\mathcal{F}}}(\frac{\log n}{n})}{k}}\right) \ a.s. \ (j=0,1,\ldots,p), \nonumber
	\end{equation}
	\begin{equation}
		\sup_{\chi\in S_{\mathcal{F}}} \left|g_{j}(\chi)-\sum_{i=1}^{n}\omega_{k}(\chi,\mathcal{X}_i)g_j(\mathcal{X}_i) \right|=O\left(\phi^{-1}\left(\frac{k}{n}\right) ^{\alpha}+\sqrt{\frac{\psi_{S_{\mathcal{F}}}(\frac{\log n}{n})}{k}}\right) \ a.s. \ (j=0,1,\ldots,p) \nonumber
	\end{equation}
	and
	\begin{equation}
		\sup_{\chi\in S_{\mathcal{F}}}\left|\sum_{i=1}^{n}\omega_{k}(\chi,\mathcal{X}_i)U_{ij} \right|= O\left(\phi^{-1}\left(\frac{k}{n}\right) ^{\alpha}+\sqrt{\frac{\psi_{S_{\mathcal{F}}}(\frac{\log n}{n})}{k}}\right) \ a.s. \ (j=1,\ldots,p), \nonumber
	\end{equation}
	where $g_0(\cdot)=m(\cdot)$ and $X_{i0}=g_0(\mathcal{X}_i)+\varepsilon_i.$ 
\end{lemma}

\begin{lemma}\label{lemma2} (\citealt{linav20}, Lemma 2) Under Assumption (A3) we have that\\
	\begin{equation}
		\max_{1\leq i,j\leq n}| \omega_{k}(\mathcal{X}_i,\mathcal{X}_j) |=O\left(\frac{1}{k}\right). \nonumber
	\end{equation}
\end{lemma}

\begin{lemma}\label{lemma3} (\citealt{linav20}, Lemma 3) Under conditions (A1)-(A5), if, in addition, $\forall r\geq 3, \ 1\leq j\leq p,$ $\mathbb{E}(|X_{1j}|^{r}|\mathcal{X}_{1}=\chi  )\leq C<\infty,$ then we have that\\
	\begin{equation}
		\frac{1}{n}\widetilde{\mathbf{X}}^{T}\widetilde{\mathbf{X}}\rightarrow \mathbf{B} \text{ a.s}. \nonumber
	\end{equation}
\end{lemma}

\begin{lemma}\label{lemma4} (\citealt{anev06}, Lemma 3) Let $V_{1},\ldots,V_{n}$ be independent r.v. with zero means and such
	that for some $r\geq2$, $\max_{1\leq j\leq n}E\left\vert V_{j}\right\vert
	^{r}\leq C<\infty$. Assume that $\left\{  a_{ij},\text{ }i,j=1,\ldots,n\right\}
	$ is a sequence of positive numbers such that $\max_{1\leq i,j\leq
		n}\left\vert a_{ij}\right\vert =O\left(  a_{n}\right)  $ and $\max_{1\leq
		i\leq n}\sum_{j=1}^{n}\left\vert a_{ij}\right\vert =O\left(  b_{n}\right)  $.
	If, in addition,
	\begin{equation}
		\exp\left(  -\dfrac{b_{n}^{1/2}\left(  \log n\right)  ^{2}%
		}{b_{n}^{1/2}+a_{n}^{1/2}n^{1/r}\log n}\right)  =O\left(  n^{-a}\right) ,
		(a>2) ,\nonumber\\
	\end{equation}
	and
	\begin{equation}
		a_{n}^{1/2}n^{1/r+b}=O\left(  b_{n}^{1/2}\log n\right),
		(b>0),\nonumber\\
	\end{equation}
	then
	\begin{equation}
		\max_{1\leq i\leq n}\left\vert \sum_{j=1}^{n}a_{ij}V_{j}\right\vert =O\left(
		a_{n}^{1/2}b_{n}^{1/2}\log n\right)  \text{ a.s.}\nonumber\\
	\end{equation}
	As a matter of fact, the conclusion of Lemma \ref{lemma4} remains unchanged
	when $\left\{  a_{ij},\text{ }i,j=1,\ldots,n\right\}  $ is a random sequence
	satisfying the conditions above almost surely.
\end{lemma}

\begin{lemma}\label{lemma4.5} Suppose that $V_1,\ldots,V_n$ are independent r.v. with zero means verifying that for some $r\geq2$, $\max_{1\leq j\leq n}E\left\vert V_{j}\right\vert
	^{r}\leq C<\infty$. In addition, suppose that assumption (A3) holds. Then\\
	\begin{equation}
		\max_{1\leq i\leq n}\left\vert \sum_{j=1}^{n}\omega_{k}(\mathcal{X}_i,\mathcal{X}_j)V_{j}\right\vert =O\left(
		k^{-1/2}\log n\right)  \text{ a.s.}\nonumber\\
	\end{equation}
\end{lemma}
\textit{Proof.} It suffices to apply Lemma \ref{lemma4} considering $a_{ij}=\omega_{k}(\mathcal{X}_i,\mathcal{X}_j)$, $a_n=1/k$ (see Lemma \ref{lemma2}) and $b_n=1$ (see (\ref{pesos})). $\Box$

\begin{lemma}\label{lemma6} Suppose that Assumption (A3) hold. If, in addition, for some $r\geq 3$ and $\forall \ 1\leq j\leq p,$ $\mathbb{E}(|\varepsilon_1|^{r})\leq C<\infty$ and $\mathbb{E}(|U_{1j}|^{r})\leq C<\infty,$ then we have that\\
	$$
	n^{-1/2}\sum_{i=1}^n\sum_{j=1}^n\omega_{k}(\mathcal{X}_i,\mathcal{X}_j)\varepsilon_j U_i=o(1) \ a.s.,
	$$
	$$
	n^{-1/2}\sum_{i=1}^n\sum_{j=1}^n\omega_{k}(\mathcal{X}_i,\mathcal{X}_j)\varepsilon_j\varepsilon_i=o(1) \ a.s.
	$$
	and
	$$
	n^{-1/2}\sum_{i=1}^n\sum_{j=1}^n\omega_{k}(\mathcal{X}_i,\mathcal{X}_j)U_jU_i^{\top}=o(1)  \ a.s.
	$$
\end{lemma}

\textit{Proof.} We only show the proof of the first result. The proofs of the others are similar.

Let us denote $a_{i}=\sum_{j=1}^n\omega_{k}(\mathcal{X}_i,\mathcal{X}_j)\varepsilon_j$; then, considering $V_j=\varepsilon_j$ in Lemma \ref{lemma4.5}, we obtain that $\max_{1\leq i\leq n}\left\vert a_{i}\right\vert =O\left(k^{-1/2}\log n\right)  \ a.s.$ Now, using that result together with Lemma \ref{lemma4} we obtain that
$$\sum_{i=1}^n a_i U_{is}=O(n^{1/2}k^{-1/2}\log^2 n)=o(n^{1/2}) \ a.s. \ (s=1,\ldots,p),$$
which concludes the proof. $\Box$

\begin{lemma}\label{lemma6.5} Under conditions (A1)-(A6)(i) we have that\\
	\begin{equation}
		\frac{1}{n}\widetilde{\mathbf{W}}^{T}\widetilde{\mathbf{W}}=\mathbf{B} + \mathbf{\Sigma}_{uu} +o(1) \ a.s., \nonumber
	\end{equation}
	\begin{equation}
		\frac{1}{n}\widetilde{\mathbf{W}}^{T}\widetilde{\mathbf{Y}} =\mathbf{B} \pmb{\beta}_0
		+o(1) \ a.s. \nonumber
	\end{equation}
	and
	\begin{equation}
		\frac{1}{n}\widetilde{\mathbf{Y}}^{T}\widetilde{\mathbf{Y}}= \pmb{\beta}_0^{\top}\mathbf{B} \pmb{\beta}_0+\sigma^2+o(1) \ a.s.
		\nonumber
	\end{equation}
\end{lemma}
\textit{Proof.} This proof is similar to that of Lemma A.7 in Liang et al. (1999). The only changes consist in considering $b_n=1/k$ instead of $b_n=n^{-4/5}$ and to apply our lemmas \ref{lemma1}, \ref{lemma3}, \ref{lemma4.5} and \ref{lemma6} instead of their lemmas A.1, A.2, A.4 and A.6, respectively, and our Lemma \ref{lemma2} instead of their Assumption 1.3(ii). $\Box$

\begin{lemma}\label{lemma7} Suppose that assumptions (A1)-(A6)(i) hold. Then\\
	\begin{equation}
		n^{-1/2}\sum_{i=1}^n\widetilde{\varepsilon}_i\widetilde{X}_i=n^{-1/2}\sum_{i=1}^n\varepsilon_i\eta_i+o(1) \ a.s., \label{uno}
	\end{equation}
	\begin{equation}
		n^{-1/2}\sum_{i=1}^n\widetilde{\varepsilon}_i\widetilde{U}_i=n^{-1/2}\sum_{i=1}^n\varepsilon_iU_i+o(1) \ a.s., \label{dos}
	\end{equation}
	\begin{equation}
		n^{-1/2}\sum_{i=1}^n\widetilde{U}_i\widetilde{U}_i^{\top}=n^{-1/2}\sum_{i=1}^nU_iU_i^{\top}+o(1) \ a.s. \label{tres}
	\end{equation}
	and
	\begin{equation}
		n^{-1/2}\sum_{i=1}^n\widetilde{X}_i\widetilde{U}_i^{\top}=n^{-1/2}\sum_{i=1}^n\eta_iU_i^{\top}+o(1) \ a.s. \label{cuatro}
	\end{equation}
\end{lemma}

\textit{Proof.} The proof of (\ref{uno}) really is part of the proof of Theorem 1 in \cite{linav20}. Specifically, they proved that
$$
S_{n3}-S_{n2}=\sum_{i=1}^n\varepsilon_i\eta_i+o(n^{1/2}) \ a.s.
$$
(for the notation $S_{n2}$ and $S_{n3}$, see (7.6) in \citealt{linav20}). In addition, from direct calculation one has that
$$
S_{n3}-S_{n2}=\sum_{i=1}^n\widetilde{\varepsilon}_i\widetilde{X}_i.
$$
(\ref{dos}) follows in a similar way as (\ref{uno}). Focusing now in (\ref{tres}), one has that
\begin{center}
	$%
	\begin{array}{lll}
		(\widetilde{\mathbf{U}}^{\top}\widetilde{\mathbf{U}})_{sl} & = & \sum_{i=1}^nU_{is}U_{il}- \sum_{i=1}^n\left(\sum_{j=1}^n\omega_{k}(\mathcal{X}_i,\mathcal{X}_j)U_{js}\right)U_{il}
		\\ 
		&  & - \sum_{i=1}^n\left(\sum_{j=1}^n\omega_{k}(\mathcal{X}_i,\mathcal{X}_j)U_{jl}\right)U_{is} \\ 
		&  & + \sum_{i=1}^n\left(\sum_{j=1}^n\omega_{k}(\mathcal{X}_i,\mathcal{X}_j)U_{js}\right)\left(\sum_{j=1}^n\omega_{k}(\mathcal{X}_i,\mathcal{X}_j)U_{jl}\right)\\ 
		& = &\sum_{i=1}^nU_{is}U_{il}+o(n^{1/2})+O(nk^{-1}\log^2n) \ a.s.
	\end{array}%
	$
\end{center}
(Note that in the last equality we have applied Lemma \ref{lemma6} and Lemma \ref{lemma4.5}). Finally, we will prove (\ref{cuatro}). We have that
\begin{equation}
	\sum_{i=1}^n\widetilde{X}_i\widetilde{U}_i^{\top}=\left(\sum_{i=1}^n\widetilde{\eta}_i\widetilde{U}_i^{\top} +
	\sum_{i=1}^n\widetilde{G}_i\widetilde{U}_i^{\top}
	\right). \label{4.1}
\end{equation}
In a similar way as in the proof of (\ref{tres}), we obtain that
\begin{equation}
	\sum_{i=1}^n\widetilde{\eta}_i\widetilde{U}_i^{\top}=\sum_{i=1}^n\eta_iU_i^{\top}+o(n^{1/2}) \ a.s.
	\label{4.2}
\end{equation}
In addition, we have that
\begin{equation}
	\left(\sum_{i=1}^n\widetilde{G}_i\widetilde{U}_i^{\top}\right)_{sl}  =  \sum_{i=1}^n \widetilde{G}_{is}U_{il}- 
	\sum_{i=1}^n\left(\sum_{j=1}^n\omega_{k}(\mathcal{X}_i,\mathcal{X}_j)U_{jl}\right)\widetilde{G}_{is}. \label{4.3}
\end{equation}%
Now, if in Lemma \ref{lemma4} we consider $a_{ij}=\widetilde{G}_{is}$, $a_n=\phi^{-1}\left(\frac{k}{n}\right) ^{\alpha}+\sqrt{\frac{\psi_{S_{\mathcal{F}}}(\frac{\log n}{n})}{k}}$ (see Lemma \ref{lemma1}) and $V_i=U_{il}$, we obtain that
\begin{eqnarray}
	\sum_{i=1}^n \widetilde{G}_{is}U_{il}&=&O_{a.s.}\left(\left( \phi^{-1}\left(\frac{k}{n}\right) ^{\alpha}+\sqrt{\frac{\psi_{S_{\mathcal{F}}}(\frac{\log n}{n})}{k}}  \right) n^{1/2} \log n \right) \nonumber \\&=&o(n^{1/2}) \ a.s. \label{4.4}
\end{eqnarray}
Finally, we have that
\begin{center}
	\begin{eqnarray}
		\left|\sum_{i=1}^n\left(\sum_{j=1}^n\omega_{k}(\mathcal{X}_i,\mathcal{X}_j)U_{jl}\right)\widetilde{G}_{is}\right| & \leq&
		n\max_i \left|\widetilde{G}_{is}\right|  \max_i\left|\left(\sum_{j=1}^n\omega_{k}(\mathcal{X}_i,\mathcal{X}_j)U_{jl}\right)\right| \nonumber \\
		& =& O\left(n\left( \phi^{-1}\left(\frac{k}{n}\right) ^{\alpha}+\sqrt{\frac{\psi_{S_{\mathcal{F}}}(\frac{\log n}{n})}{k}}  \right)\right) \nonumber \\
		& \times& O\left( \log n/\sqrt{k} \right) \nonumber \\
		&=& o(n^{1/2}) \ a.s. \label{4.5} 
	\end{eqnarray}
	
\end{center}
(Note that in the first equality above we have used lemmas \ref{lemma1} and \ref{lemma4.5}) (\ref{4.1})-(\ref{4.5}) conclude the proof. $\Box$

\begin{lemma}\label{lemmafin} Suppose that assumptions (A1)-(A6)(i). Then\\
	\begin{equation}
		\sum_{i=1}^n\widetilde{U}_i\widetilde{m}_i=o(n^{1/2}) \ a.s. \text{ and }
		\sum_{i=1}^n\widetilde{X}_i\widetilde{m}_i=o(n^{1/2}) \ a.s. 
		\nonumber
	\end{equation}

\end{lemma}
\textit{Proof.} The first result can be obtained from (\ref{4.4}) and (\ref{4.5}) by considering $\pmb{m}$ instead of $\mathbf{G}$. Focusing now on the second result, we have that 
\begin{equation}
	\sum_{i=1}^n\widetilde{X}_i\widetilde{m}_i=
	\sum_{i=1}^n\widetilde{G}_i\widetilde{m}_i+
	\sum_{i=1}^n\widetilde{\eta}_i\widetilde{m}_i. \label{4.6}
\end{equation}
From Lemma \ref{lemma1} one obtains that first summation in the right hand of equality (\ref{4.6}) is $o(n^{1/2}) \ a.s$. Finally, the order of the second summation, $o(n^{1/2}) \ a.s.$, can be obtained in a similar way as that of the first result in this lemma (considering $\eta_i$ instead of $U_i$). $\Box$

\subsection{Proof of Theorem \ref{theorem1}}\label{app2}

\textit{Proof of Theorem \ref{theorem1}(i).}

Let us denote $\mathbf{M}_n= (\widetilde{\mathbf{W}}^{\top}\widetilde{\mathbf{W}}-n\pmb{\Sigma}_{uu})/n$. From (\ref{beta-est}) together with the facts that $\widetilde{\mathbf{W}}= \widetilde{\mathbf{X}} + \widetilde{\mathbf{U}}$ and $\widetilde{\mathbf{Y}}= \widetilde{\mathbf{X}}\pmb{\beta}_0 + \widetilde{\mathbf{m}} + \widetilde{\pmb{\varepsilon}}$, we can write
\begin{eqnarray}
	n^{1/2}(\widehat{\pmb{\beta}}_{0k}-\pmb{\beta}_0)&=&n^{-1/2}\mathbf{M}_n^{-1}\left(\widetilde{\mathbf{X}}^{\top}\widetilde{\mathbf{m}} + \widetilde{\mathbf{X}}^{\top}\widetilde{\pmb{\varepsilon}} + \widetilde{\mathbf{U}}^{\top}\widetilde{\mathbf{m}} + \widetilde{\mathbf{U}}^{\top}\widetilde{\pmb{\varepsilon}}\right. \nonumber \\ && \left.- \widetilde{\mathbf{X}}^{\top}\widetilde{\mathbf{U}}\pmb{\beta}_0 - \widetilde{\mathbf{U}}^{\top}\widetilde{\mathbf{U}}\pmb{\beta}_0 +n\pmb{\Sigma}_{uu}\pmb{\beta}_0\right). \label{norm}
\end{eqnarray}
Then, taking (\ref{norm}) into account, from a direct application of our lemmas \ref{lemma6.5}
, \ref{lemma7} and \ref{lemmafin} we obtain that
\begin{equation}
	n^{1/2}(\widehat{\pmb{\beta}}_{0k}-\pmb{\beta}_0)=n^{-1/2}\mathbf{B}^{-1}\sum_{i=1}^n\left(\varepsilon_i\eta_i +\varepsilon_iU_i -\eta_iU_i^{\top}\pmb{\beta}_0 -U_iU_i^{\top}\pmb{\beta}_0 + \pmb{\Sigma}_{uu}\pmb{\beta}_0 \right)+o(1) \ a.s. \label{norm2}
\end{equation}
Let us denote 
\begin{equation}
	\zeta_i=\varepsilon_i\eta_i +\varepsilon_iU_i -\eta_iU_i^{\top}\pmb{\beta}_0 -U_iU_i^{\top}\pmb{\beta}_0 + \pmb{\Sigma}_{uu}\pmb{\beta}_0. \label{norm3}
\end{equation}
To finish the proof of this theorem, it remains to prove that the sequence of random variables $\{\zeta_i\}$ verifies the Lindeberg condition, and that
\begin{equation}
	n^{-1}\sum_{i=1}^n \text{cov}(\zeta_i)= \pmb{\varGamma}+o(1). \label{norm4}
\end{equation}
(For the definition of $\pmb{\varGamma}$, see Theorem \ref{theorem1}) Those two conditions can be verified in the same way as in Liang et al. (1999), pages 1533-1534; so, for the sake of brevity, such proofs are omitted here. Note that, although in the model of Liang et al. (1999) there are no functional variables and the estimation procedure (kerned-based) is different from the one here (\textit{k}NN-based), such facts do not affect to $\zeta_i$. $\Box$

\textit{Proof of Theorem \ref{theorem1}(ii).}

From (\ref{norm2}), we can write
\begin{equation}
	\widehat{\pmb{\beta}}_{0k}-\pmb{\beta}_0=\mathbf{B}^{-1}n^{-1}\sum_{i=1}^n \zeta_i +o(n^{-1/2}) \ a.s., \label{L}
\end{equation}
where the random vector $\zeta_i$ was defined in (\ref{norm3}). 

Let $\mathbf{b}_j=(b_{j1}, \ldots, b_{jp})^{\top}$ denote the $j$-th row of the matrix $\mathbf{B}^{-1}$. Considering $V_i=\mathbf{b}_j^{\top}\zeta_i/Var(\mathbf{b}_j^{\top}\zeta_i)^{1/2}$ in Lemma \ref{LemaStout} and taking (\ref{norm4}) into account, we obtain that
\begin{equation}
	\limsup_{n\rightarrow \infty}\left(\frac{1}{2n\log \log n}\right)^{1/2}\left|\sum_{i=1}^n  \mathbf{b}_j^{\top}\zeta_i\right|=\sigma_{jj}^{1/2} \ a.s. \label{LL}
\end{equation}
(for the definition of $\sigma_{jj}$, see Theorem \ref{theorem1}(ii)). Finally, the claimed result follows from (\ref{L})-(\ref{LL}). $\Box$

\textit{Proof of Theorem \ref{theorem1}(iii).}

The main idea of the proof consists in applying existing results for $k$NN estimators in the functional nonparametric regression model without additional multivariate predictors, and then to deal with the question of estimating the additional linear coefficients $\pmb{\beta}_0$. For fixed $\chi\in\mathcal{H}$, we have that:
\begin{eqnarray}
	\left|\widehat{m}_{k}(\chi)-m(\chi)\right|
	\leq \left|\sum_{i=1}^n \omega_{k}(\chi,\mathcal{X}_i)\left(m(\mathcal{X}_i)+\varepsilon_i\right)-m(\chi)\right| \nonumber \\
	+\left|\sum_{j=1}^p \widehat{g}_{j,k}(\chi)\left(\beta_{0j}-\widehat{\beta}_{0kj}\right)\right| 
	+\left|\sum_{j=1}^p\sum_{i=1}^n \omega_{k}(\chi,\mathcal{X}_i)U_{ij}\widehat{\beta}_{0kj}\right| \nonumber \\
	\equiv A_1(\chi)+A_2(\chi)+A_3(\chi),\label{PKNN11}
\end{eqnarray}
where $\widehat{g}_{j,k}(\chi)$ denotes the $k$NN estimator of $g_{j}(\chi)$ $(j=1,\dots,p)$; that is, $\widehat{g}_{j,k}(\chi)=\sum_{i=1}^n \omega_{k}(\chi,\mathcal{X}_i)X_{ij}.$ Considering $m(\chi)$ and $m(\mathcal{X}_i)+\varepsilon_i$ instead of $g_j(\chi)$ and $X_{ij}$, respectively, in Lemma \ref{lemma1}, we obtain that
\begin{equation}
	\sup_{\chi\in S_{\mathcal{F}}}A_1(\chi)= O\left(\phi^{-1}\left(\frac{k}{n}\right) ^{\alpha}+\sqrt{\frac{\psi_{S_{\mathcal{F}}}(\frac{\log n}{n})}{k}}\right) \ a.s.
	\label{PKNN21}
\end{equation}
In addition, applying Lemma \ref{lemma1} again together with Theorem \ref{theorem1}(ii), we have that
\begin{equation}
	\sup_{\chi\in S_{\mathcal{F}}}A_2(\chi)= O\left(\frac{\log \log n}{n}\right)^{\frac{1}{2}} \ a.s.
	\label{PKNN22}
\end{equation}
Finally, 
Lemma \ref{lemma1} together with Theorem \ref{theorem1}(ii) give
\begin{equation}
	\sup_{\chi\in S_{\mathcal{F}}}A_3(\chi)= O\left(\phi^{-1}\left(\frac{k}{n}\right) ^{\alpha}+\sqrt{\frac{\psi_{S_{\mathcal{F}}}(\frac{\log n}{n})}{k}}\right) \ a.s.
	\label{PKNN23}
\end{equation}
\noindent The claimed result is obtained from (\ref{PKNN21})-(\ref{PKNN23}). $\Box$

\end{document}